\documentclass[aps,prl,reprint,letterpaper,groupedaddress,floatfix,superscriptaddress,longbibliography,nobibnotes]{revtex4-2}

\usepackage{bm} 
\usepackage{amsmath}
\usepackage{amssymb}
\usepackage{mathtools}
\usepackage{braket}
\usepackage{ascmac}
\usepackage{siunitx}
\usepackage{natbib}
\usepackage{multirow}
\usepackage{cancel}

\usepackage{hyperref}
\usepackage{color}
\usepackage{graphicx}

\usepackage{comment}
\usepackage{dcolumn}
\usepackage{threeparttable}
\usepackage{autobreak}
\DeclareSIUnit{\cps}{cps}

\newcommand{\vack}[1]{\ket{\cancel{0}}_{#1}}

\newcommand{\cre}[1]{\hat{a}_{#1} ^{\dagger}}

\begin{document}


\title{Programmable non-Gaussian quantum light source\\with state and temporal-waveform tunability}


\author{Hiroko Tomoda}

\affiliation{Department of Applied Physics, School of Engineering, The University of Tokyo, 7-3-1 Hongo, Bunkyo-ku, Tokyo 113-8656, Japan}

\author{Yu Nishizawa}
\affiliation{Department of Applied Physics, School of Engineering, The University of Tokyo, 7-3-1 Hongo, Bunkyo-ku, Tokyo 113-8656, Japan}

\author{Akihiro Machinaga}
\affiliation{Department of Applied Physics, School of Engineering, The University of Tokyo, 7-3-1 Hongo, Bunkyo-ku, Tokyo 113-8656, Japan}

\author{Takahiro Kashiwazaki}
\affiliation{Device Technology Labs, NTT, Inc., 3-1, Morinosato Wakamiya, Atsugi, Kanagawa 243-0198, Japan}

\author{Takeshi Umeki}
\affiliation{Device Technology Labs, NTT, Inc., 3-1, Morinosato Wakamiya, Atsugi, Kanagawa 243-0198, Japan}

\author{Shigehito Miki}
\affiliation{Advanced ICT Research Institute, National Institute of Information and Communications Technology, 588-2 Iwaoka, Nishi-ku, Kobe, Hyogo 651-2492, Japan}

\author{Masahiro Yabuno}
\affiliation{Advanced ICT Research Institute, National Institute of Information and Communications Technology, 588-2 Iwaoka, Nishi-ku, Kobe, Hyogo 651-2492, Japan}

\author{Hirotaka Terai}
\affiliation{Advanced ICT Research Institute, National Institute of Information and Communications Technology, 588-2 Iwaoka, Nishi-ku, Kobe, Hyogo 651-2492, Japan}

\author{Daichi Okuno}
\affiliation{Department of Applied Physics, School of Engineering, The University of Tokyo, 7-3-1 Hongo, Bunkyo-ku, Tokyo 113-8656, Japan}

\author{Shuntaro Takeda}
\email{takeda@ap.t.u-tokyo.ac.jp}
\affiliation{Department of Applied Physics, School of Engineering, The University of Tokyo, 7-3-1 Hongo, Bunkyo-ku, Tokyo 113-8656, Japan}


\date{\today}

\begin{abstract}
A versatile quantum light source capable of programmably generating a variety of quantum light is a key enabler for photonic quantum technologies.
In particular, independent control over both the output quantum state and its temporal waveform is essential for realizing diverse functionalities and enhancing processing performance.
However, conventional sources of optical non-Gaussian states, a crucial resource for photonic quantum information processing, typically emit fixed states with predetermined temporal waveforms, lacking their programmability.
Here, we propose a programmable non-Gaussian quantum light source that offers independent and arbitrary tunability of both the quantum state and the temporal waveform within a single platform.
As a distinctive feature, our approach employs a heralding scheme in which these two properties are indirectly engineered to user-defined targets by manipulating the light in the heralding channel, thereby avoiding optical losses associated with direct manipulation of the heralded quantum light.
We develop a prototype and demonstrate the generation of single-photon, Schr\"odinger cat, and two-photon states in a variety of unconventional temporal waveforms without degradation in state quality.
This platform provides a versatile tool for tailoring quantum light to specific applications, significantly expanding the capabilities of photonic quantum technologies.
\end{abstract}


\maketitle



The advancement of photonic quantum technologies~\cite{Flamini2019Photonic,Takeda2019large} goes hand in hand with progress in quantum light sources.
To date, the pursuit of photonic quantum computing, communication, and sensing has driven the development of a wide variety of quantum light sources, including single-photon sources~\cite{Eisaman2011Invited}, squeezed-light sources~\cite{Andersen201630}, and more advanced non-Gaussian light sources~\cite{Lvovsky2020Production} capable of generating higher-order Fock states~\cite{Cooper2013Experimental,Sonoyama2024Generation} and Schr\"odinger cat states~\cite{Ourjoumtsev2006Generating,Takahashi2008Generation}.
Since each of these sources is typically engineered to produce a specific type of quantum light, different applications generally require different light sources.

A versatile quantum light source capable of programmably generating a variety of quantum light within a single device would therefore be a key enabler for the next stage of photonic quantum technologies.
In particular, programmability over both the output quantum state and the temporal waveform (TW) is indispensable for realizing such versatility.
Programmability over the quantum state allows multiple quantum protocols to be executed on the same hardware, accelerating the development and deployment of diverse applications.
This functionality also enables state optimization under realistic noisy conditions, thereby maximizing the performance of quantum protocols, as exemplified by variational-state quantum metrology~\cite{Koczor2020Variational}.
Furthermore, the ability to programmably shape the TWs provides additional benefits, including enhanced processing fidelity by optimizing temporal-mode matching, efficient coupling to matter-based quantum memories via exponentially rising waveforms~\cite{Stobinska2009Perfect,Wang2011Efficient}, mitigation of low-frequency noise via DC-free waveforms~\cite{Yoshikawa2016Invited,Takeda2012Quantum}, and higher information capacity through temporal encoding~\cite{Brecht2015Photon}.
Thus, programmability over both the state and the TW would significantly expand the capabilities of photonic quantum technologies.

\begin{figure}[!b]
 \centering
 \includegraphics[trim={0mm 0mm 12pt 0mm},clip,width=\linewidth]{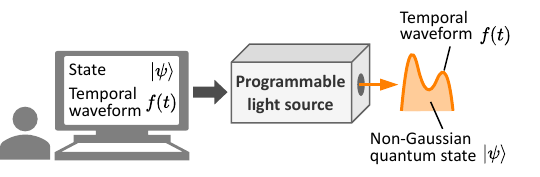}
 \caption{
   Schematic image of programmable non-Gaussian quantum light source with state and temporal-waveform tunability.
   \label{fig:Fig1}
   }
\end{figure}

However, conventional quantum light sources generally emit fixed quantum states with predetermined TWs, lacking programmability.
This limitation arises from the fragility of quantum light; direct manipulation to modify its state or TW often introduces optical losses, leading to the degradation of its nonclassical features.
While programmable control is relatively straightforward for Gaussian quantum states owing to their simple generation mechanism~\cite{Tomoda2023Programmable}, it remains challenging for non-Gaussian states, although such states are crucial for unlocking the full potential of quantum technologies~\cite{Bartlett2002Efficient,Niset2009No,Eisert2002Distilling}.
Previous studies have demonstrated programmability in either the quantum state or the TW, but simultaneous and independent control of both remains to be realized.
For example, some non-Gaussian light sources allow state control by tuning system parameters~\cite{Bimbard2010Quantum,Yukawa2013Generating,Larsen2025Integrated,Takase2026Tuning}.
Meanwhile, various techniques for programmably shaping TWs have been demonstrated for single-photon states~\cite{Keller2004Continuous,Farrera2016Generation,Ansari2018Heralded}, but these are not readily applicable to more general non-Gaussian states.
Although another methodology capable of arbitrary TW shaping in general non-Gaussian state generation was proposed in Ref.~\cite{Takase2022Quantum}, it requires physical reconfiguration of the optical setup for each target waveform, thereby limiting its practical programmability.

Here, we propose a programmable non-Gaussian quantum light source that enables independent and arbitrary control of both the quantum state and the TW within a single optical setup (Fig.~\ref{fig:Fig1}).
Our approach employs a heralding scheme in which TW- and state-engineering modules are implemented in the heralding channel, by extending a previously proposed (but not experimentally demonstrated) method for TW shaping of single photons~\cite{Averchenko2017Temporal}.
This configuration allows for indirect manipulation of the generated quantum light without compromising its quality.
Furthermore, this light source allows the generation timing to be easily synchronized with other quantum devices, overcoming a long-standing bottleneck in continuous-wave (CW)-based non-Gaussian light sources.
We develop a prototype of this system and demonstrate the generation of single-photon, Schr\"odinger cat, and two-photon states in various unconventional TWs without hardware reconfiguration.
Comparison with a conventional non-programmable scheme shows that our scheme can independently control both the non-Gaussian state and its TW without any degradation in nonclassical features.

Our programmable quantum light source will serve as a universal platform for a wide range of photonic quantum information processing tasks, including both qubit-based~\cite{Flamini2019Photonic} and continuous-variable-based~\cite{Takeda2019large} approaches, thereby enabling the flexible development of diverse applications.
Furthermore, by optimally tailoring quantum light to specific purposes, it provides versatile functionalities and enhanced processing performance, accelerating the advancement and precision of photonic quantum technologies.

\begin{figure*}[bthp]
 \centering
 \includegraphics[trim={0mm 0mm 63pt 0mm},clip,width=0.92\linewidth]{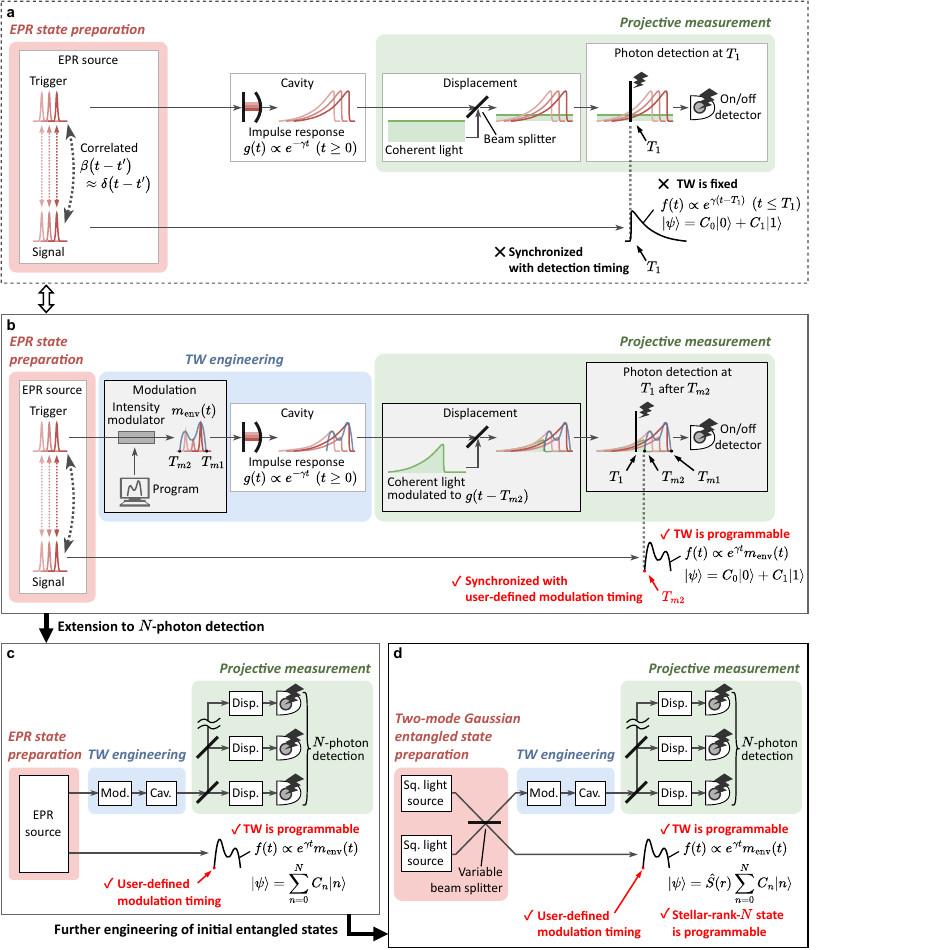}
 \caption{
   Comparison between the conventional and proposed quantum light source schemes.
   \textbf{a,}~Conventional setup for the generation of $C_0 \ket{0} + C_1 \ket{1}$,
   \textbf{b--d,}~Our proposed setups for the generation of $C_0 \ket{0} + C_1 \ket{1}$, $\sum_{n=0} ^{N} C_n \ket{n}$, and $\hat{S}(r) \sum_{n=0} ^{N} C_n \ket{n}$, respectively.
   EPR state, Einstein-Podolsky-Rosen state;
   TW, temporal waveform;
   Mod., modulation;
   Cav., cavity;
   Disp., displacement;
   Sq. light, squeezed light.
   From \textbf{a} to \textbf{b}, we introduce a trigger modulation system before the cavity, modulated coherent light for implementing displacement, and a time window ($T_1\geq T_{m2}$) for photon detection.
   From \textbf{b} to \textbf{c}, we extend the projective measurement system to realize $N$-photon detection, thereby generating $\sum_{n=0} ^{N} C_n \ket{n}$.
   From \textbf{c} to \textbf{d}, we further engineer the initial two-mode entangled states, thereby generating $\hat{S}(r) \sum_{n=0} ^{N} C_n \ket{n}$.
   \label{fig:Fig2}
   }
\end{figure*}

\section{Results}
\subsection{Working principle\label{sec:theory}}

The proposed programmable non-Gaussian quantum light source is shown in Fig.~\ref{fig:Fig2}d.
By conditioning on $N$-photon detection, this source can programmably generate any quantum state of the form
\begin{gather}
  \ket{\psi} =\hat{S}(r) \sum_{n=0}^{N} C_n \ket{n}
\label{eq:FinalState}
\end{gather}
in an arbitrary real-valued TW, where $\ket{n}$ denotes an $n$-photon Fock state, $\hat{S}(r)$ is the squeezing operator ($r \in \mathbb{R}$), and $C_0,C_1,\dots,C_N \in \mathbb{C}$ are complex superposition coefficients.
In other words, this setup can generate an arbitrary quantum state with stellar rank $N$ (a commonly used measure of non-Gaussianity~\cite{Chabaud2020Stellar}), using the minimum number of single-photon detectors, up to readily implementable displacement and phase-shift operations.
In the following, we explain the principle of our scheme step by step (the full mathematical details are provided in the Supplementary Information).
We first review the conventional scheme for generating a single-photon state and a superposition state of the form $C_0 \ket{0} + C_1 \ket{1}$ with a fixed TW (Fig.~\ref{fig:Fig2}a).
We then propose a programmable TW-shaping scheme for these states using a modified setup (Fig.~\ref{fig:Fig2}b), and finally extend it to arbitrary states of the form given in Eq.~\eqref{eq:FinalState} (Figs.~\ref{fig:Fig2}c and d).

First, we describe the conventional single-photon generation scheme shown in Fig.~\ref{fig:Fig2}a~\cite{Takase2022Generation,Tomoda2024Boosting}.
Here, an initial Einstein-Podolsky-Rosen (EPR) state in the trigger and signal channels is assumed to be generated continuously, for example by an optical parametric amplifier (OPA) pumped by CW light.
The trigger light then passes through a filtering cavity.
Finally, single-photon detection in the trigger channel heralds the preparation of a single-photon state in the signal channel.
When multiphoton contributions are negligible, on/off detectors can be employed for this detection.
In this scheme, the TW $f(t)$ of the heralded state is determined by the cavity's response.
This mechanism can be understood as follows.
As shown in Fig.~\ref{fig:Fig2}a, the cavity mixes the trigger light at different times according to its response function $g(t)\propto e^{-\gamma t}\ (t\geq 0)$ before photon detection.
When the temporal resolution of the photon detector is much shorter than the cavity decay constant $1/\gamma$, a photon detection at time $T_1$ can be regarded as a projection of the trigger light onto $\cre{\text{t}}(T_1)\vack{\text{t}}$.
Here we denote the photon-creation operator at time $t$ and the multimode vacuum state in Channel $i$ as $\cre{i}(t)$ and $\vack{i}$, respectively.
Such a photon detection is equivalent to projecting the trigger light prior to the cavity onto $\ket{\phi}_{\text{t}}=\int dt g(T_1-t) \cre{\text{t}}(t) \vack{\text{t}}$, namely a single-photon state with TW $g(T_1-t)$.
Assuming that the photon pair in the signal and trigger channels is described by $\ket{\Phi}_{\text{s,t}}=\iint dt dt' \beta(t-t') \cre{\text{s}}(t) \cre{\text{t}}(t') \vack{\text{s}} \vack{\text{t}}$~\cite{Loudon2000Quantum}, and that the temporal width of its time-correlation function $\beta(t-t')$ is sufficiently shorter than the cavity decay constant, i.e., $\beta(t-t') \approx \delta(t-t')$, the heralded signal photon is calculated as
\begin{gather}
{}_\text{t}\braket{\phi|\Phi}_{\text{s,t}}=\int dt g(T_1-t) \cre{\text{s}}(t) \vack{\text{s}}.
\end{gather}
Thus the TW $f(t)$ of the heralded photon is determined by the cavity response function $g(t)$ as
\begin{gather}
  f(t)=g(T_1-t) \propto e^{\gamma(t-T_1)} \quad (t\leq T_1),
\label{eq:f_inConventional}
\end{gather}
and the generation timing is synchronized with the detection time $T_1$.
Intuitively, this can be understood as a consequence of the strong temporal correlation between the trigger and signal photons, which causes their TWs to become identical: the cavity selects the projection TW of the trigger photon, which is directly mapped onto the TW of the heralded signal photon.
While the above discussion focuses on the generation of single-photon states, arbitrary superpositions of vacuum and single-photon states $C_0\ket{0}+C_1\ket{1}$ can also be generated by adding a displacement operation through the injection of coherent light with constant power into the trigger channel~\cite{Bimbard2010Quantum,Yukawa2013Generating}, as shown in Fig.~\ref{fig:Fig2}a.
The TW of such superposition states is also determined in the same way as in the single-photon case.
To summarize, in the conventional scheme, the TW is determined by the response of the cavity (or, more generally, an optical filter) in the trigger path.
Therefore, appropriate passive filters have been constructed to engineer various TWs in previous experiments~\cite{Takase2022Quantum,Takeda2012Quantum,Ogawa2016Real}.
However, their design freedom is inherently limited, rendering arbitrary TW shaping experimentally challenging.
Moreover, such passive filters do not offer programmability for optimizing the TW for specific purposes or modifying it for different applications.

As a first step to overcome such a limitation, we propose a programmable TW-shaping scheme for the state $C_0\ket{0} + C_1\ket{1}$, shown in Fig.~\ref{fig:Fig2}b, by extending a prior proposal on the TW shaping of single-photon states~\cite{Averchenko2017Temporal}.
For this purpose, we introduce an intensity modulator in the trigger path, thereby carving pulses from the continuous trigger light before it reaches the cavity and the detector.
By modulating the trigger-light amplitude with a temporal envelope $m_{\text{env}}(t)$ (a real-valued function defined in $T_{m1} \le t \le T_{m2}$) and selecting photon detection events with $T_1 \ge T_{m2}$, these detection events project the trigger light prior to the intensity modulator onto $\int dt e^{\gamma t} m_{\text{env}}(t) \cre{\text{t}}(t) \vack{\text{t}}$ in the absence of the displacement.
Due to the strong temporal correlation between the trigger and signal photons, the effect of the trigger modulation is mapped onto the signal photon. Thus, the TW of the heralded signal photon is given by
\begin{gather}
f(t) \propto e^{\gamma t} m_{\text{env}}(t),
\label{eq:f_inourscheme}
\end{gather}
and can be shaped through the modulation $m_{\text{env}}(t)$.
Since Eq.~(\ref{eq:f_inourscheme}) is independent of $T_1$,
the generation timing is synchronized with the modulation timing rather than with the photon-detection time $T_1$.
This scheme can also be extended to enable programmable generation of arbitrary superpositions $C_0\ket{0} + C_1\ket{1}$ by incorporating a displacement operation through the injection of coherent light into the trigger channel.
In this case, the coherent light, previously injected at constant power, should be modulated to match the temporal profile $g(t-T_{m2})$ of the cavity response, as shown in Fig.~\ref{fig:Fig2}b.
This ensures that the superposition coefficients $C_0$ and $C_1$ of the heralded states become identical for any detection time $T_1$ with $T_1\geq T_{m2}$.
Consequently, the TW $f(t)$ of an arbitrary superposition $C_0\ket{0} + C_1\ket{1}$ can be programmed into an arbitrary real-valued function through appropriate modulation $m_{\text{env}}(t)$, provided that the temporal width of $f(t)$ is much longer than the photon-pair correlation time and the detector's temporal resolution while remaining comparable to or shorter than the cavity decay constant $1/\gamma$.
Note that the above scheme requires the filter response to have an exponentially decaying profile (see Supplementary Information for details).
In addition, although we have so far assumed a continuous EPR source, our scheme is also applicable to a pulsed source as long as the aforementioned timescale relationship is satisfied.

\begin{figure*}[tbhp]
 \centering
 \includegraphics[width=\linewidth]{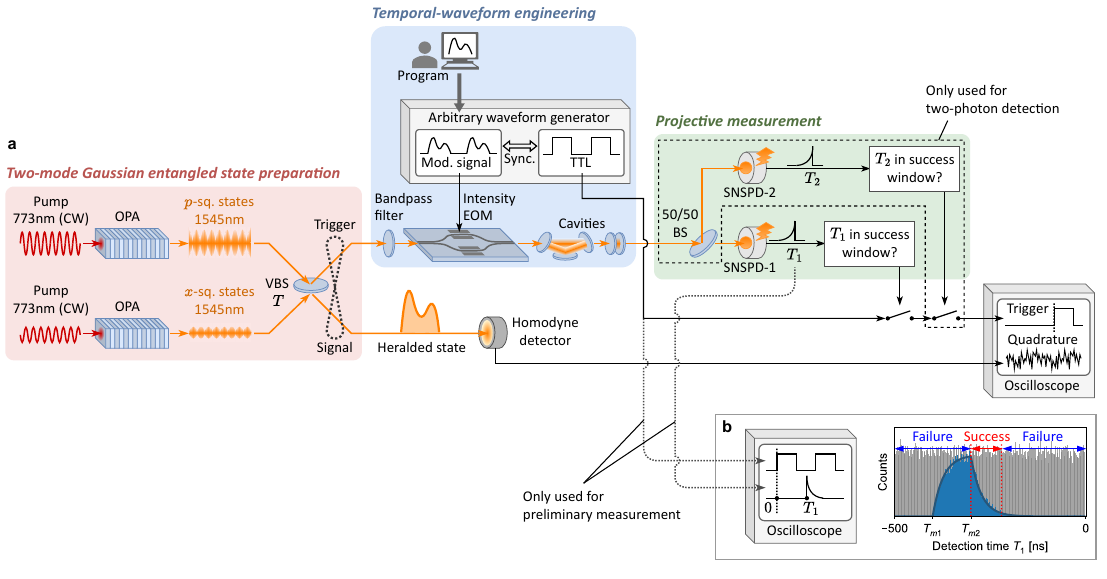}
 \caption{
   \textbf{a,~}Schematic of the experimental setup.
   CW, continuous wave;
   OPA, optical parametric amplifier;
   sq. states, squeezed states; 
   BS, beam splitter;
   VBS, variable beam splitter;
   EOM, electro-optic modulator;
   Mod. signal, modulation signal;
   Sync., synchronization;
   SNSPD, superconducting nanostrip single-photon detector.
   \textbf{b,~}Preliminary measurement for determining the success time window.
   Photon detection signals are acquired by using the TTL signal from the AWG as a trigger.
   Distributions of the photon-detection time $T_1$ are plotted as a blue histogram for our scheme with square-pulse modulation and as a gray histogram for the conventional scheme.
   The blue histogram is fitted with the theoretically predicted curve, allowing us to determine the modulation completion time, denoted as $T_{m2}$ in the theoretical section.
   Based on this result, the time window for successful temporal-waveform shaping is appropriately set.
   \label{fig:Fig3}
 }
\end{figure*}

\begin{figure*}[htbp]
 \centering
 \includegraphics[width=\linewidth]{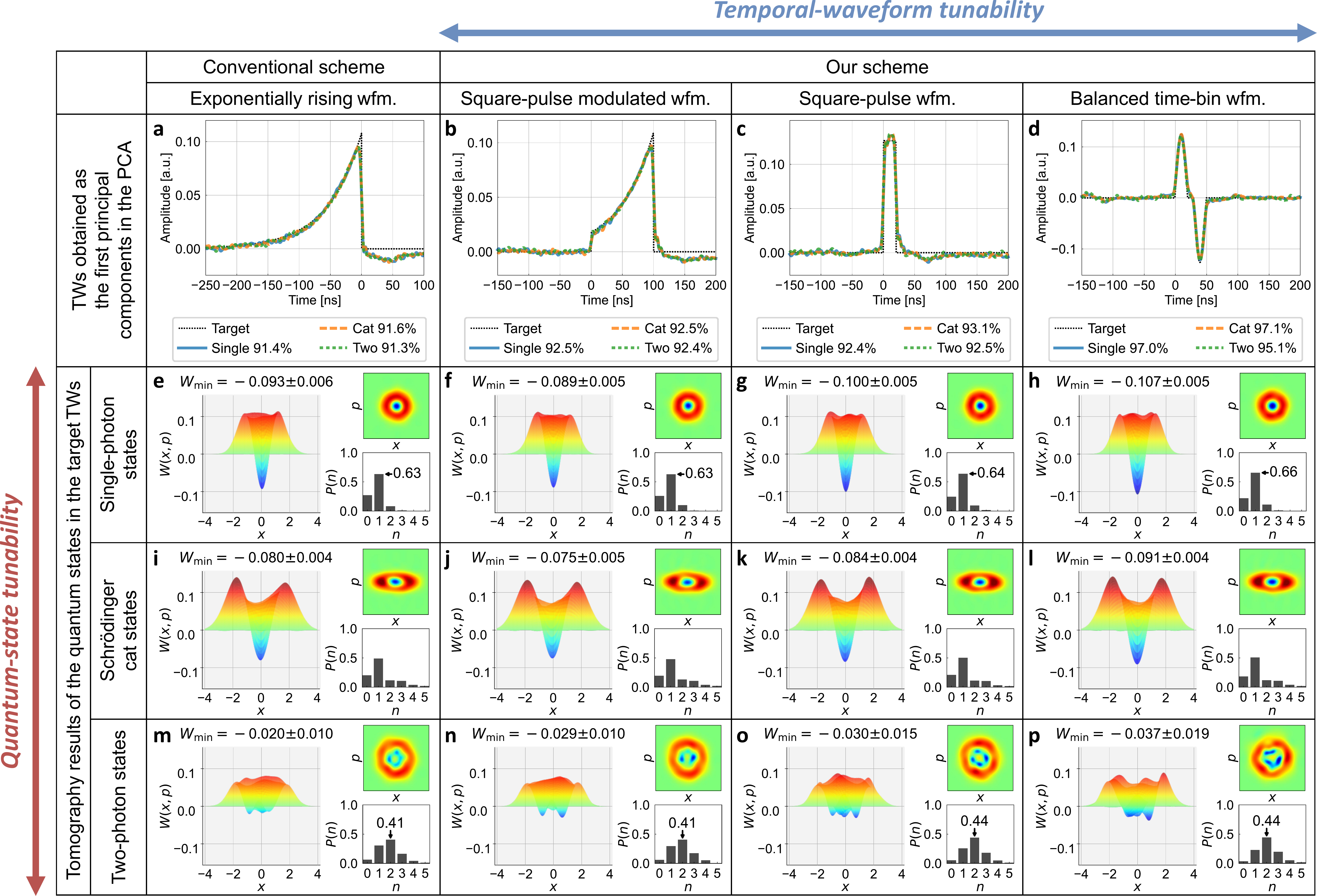}
 \caption{
   Experimental results.
   \textbf{a--d,~}
   Temporal waveforms (TWs) of the generated states.
    (wfm. indicates waveform.)
   These are obtained as the first principal components in principal component analysis (PCA).
   In each panel, the results for three types of quantum states are plotted together.
   The values shown in the legends represent the temporal-mode matching between each obtained TW and the corresponding target TW.
   \textbf{e--p,~}
   Wigner functions and photon-number distributions of the generated states ($\hbar=1$).
   These are obtained for the target TWs (black dotted lines in \textbf{a--d}).
   The minimum values of the Wigner functions are indicated with error bars, which are calculated using the Fisher information matrix~\cite{Rehacek2008Tomography}.
   \label{fig:Fig4}
   }
 \end{figure*}

We then extend the above TW-shaping scheme to general superposition states $\sum_{n=0}^{N} C_n \ket{n}$ by introducing the setup shown in Fig.~\ref{fig:Fig2}c.
The trigger path is split into $N$ channels after the cavity, with each channel including a displacement via modulated coherent light and a single on/off detector for photon detection.
This configuration projects the trigger light onto superposition states containing up to $N$ photons, and the coefficients $\{C_n\}_{n=0}^{N}$ of the final heralded states can be programmed by adjusting the intensity and phase of each displacement.

Finally, we further extend the setup by replacing the EPR state with a more general two-mode Gaussian entangled state produced by interfering two orthogonally squeezed states, as shown in Fig.~\ref{fig:Fig2}d.
This extension enables the programmable generation of $\hat{S}(r)\sum_{n=0}^{N} C_n \ket{n}$ in Eq.~\eqref{eq:FinalState}, where the output squeezing parameter $r$ is tunable through the two input squeezing levels and the beam-splitter (BS) transmissivity.
Furthermore, the tunability of the initial two-mode Gaussian entangled state can increase the generation rate, as demonstrated in our previous study~\cite{Tomoda2024Boosting}.
Consequently, our proposed setup enables simultaneous programmability of both the TW $f(t)$ and the quantum state $\ket{\psi}$.

Additionally, our scheme offers several advantageous features.
First, it indirectly reshapes the TW via trigger-light modulation, thereby avoiding additional optical losses compared with direct modulation of the signal light.
Second, it is compatible with practical technologies, as it only requires on/off detectors and does not rely on high-efficiency photon-number-resolving detectors.
Third, the generation timing of the heralded states can be easily synchronized with other quantum devices, overcoming a long-standing limitation of CW-laser-based non-Gaussian light sources. While CW-laser-based systems are advantageous in terms of higher homodyne detection efficiencies compared to their pulsed-laser-based counterparts, their main bottleneck has been the \textit{continuous and random} timing of non-Gaussian state generation, which makes it difficult to synchronize multiple such states~\cite{Konno2024Logical,Makino2016Synchronization}.
In our scheme, although the state generation itself is inherently probabilistic as the conventional scheme, generation timing is always synchronized with the user-defined modulation timing, making it easier to interface the generated states with other devices.

\subsection{Experimental verification}

As a prototype of the proposed programmable light source in Fig.~\ref{fig:Fig2}d, we construct the experimental system shown in Fig.~\ref{fig:Fig3}, which includes up to two photon-detection channels without displacement operations.

As a first step, we verify the system operation under the simplest experimental condition.
More specifically, we configure the system for single-photon generation by preparing two orthogonally squeezed states with identical squeezing levels, setting the BS transmissivity to 0.5, and detecting a single photon in the trigger path.
For TW shaping, we apply a \SI{100}{\nano\second}-wide square-pulse intensity modulation.

As described in the previous section, our scheme generates non-Gaussian states in the desired TW only when the photon-detection time $T_1$ satisfies $T_1 \geq T_{m2}$.
To properly define the success window, we first determine the modulation completion time $T_{m2}$ through a preliminary measurement in which the time distribution of photon detection events is recorded using a trigger synchronized with the modulation signal.
The result in Fig.~\ref{fig:Fig3}b shows that the detection time is uniformly distributed without modulation (gray histogram), whereas it becomes localized under modulation (blue histogram).
The time $T_{m2}$ is determined by fitting the latter distribution with a theoretical curve derived from the modulation pattern and the cavity response (blue line). The success time window in the following TW-shaping experiments is then set from the determined $T_{m2}$ to an appropriate duration (several tens of nanoseconds; see Supplementary Information).

We then proceed to evaluate the TW of the generated single-photon states under this condition.
For this purpose, we perform principal component analysis (PCA)~\cite{Morin2013Experimentally} on the acquired quadrature amplitude data and extract the TW.
In the conventional case without modulation, we obtain an exponentially rising waveform synchronized with the photon-detection timing (blue solid line in Fig.~\ref{fig:Fig4}a), as expected from Eq.~\eqref{eq:f_inConventional}.
In contrast, with the square-pulse modulation, we obtain a different TW synchronized with the modulation timing (blue solid line in Fig.~\ref{fig:Fig4}b), and this TW can be well explained by the theoretical prediction in Eq.~\eqref{eq:f_inourscheme}.
Quantitatively, temporal-mode matching between the obtained TWs and their corresponding theoretical TWs (black dotted line in each figure) is calculated to be 91.4\% for the conventional case and 92.5\% for the modulation case.
These comparable values indicate that our system can reshape the TW with high precision.

Finally, we perform quantum tomography for these theoretical TWs to compare the qualities of the generated single-photon states between the cases with and without modulation.
Figures~\ref{fig:Fig4}e and f show that the generated states exhibit comparable quality, as reflected in the similarity of their single-photon components $P(1)$, as well as the minimum values $W_{\text{min}}$ of the Wigner functions $W(x,p)$, a commonly used indicator of nonclassicality.
These results show that our TW-shaping scheme does not deteriorate the state quality, as theoretically predicted.

As the next step,
we demonstrate that our system can program the TW of a single photon into a target shape $f(t)$ by adjusting the modulation envelope $m_{\text{env}}(t)$.
We choose two target TWs $f(t)$: a 20-ns-wide square pulse as a shorter waveform, and a balanced time-bin waveform with negative amplitude, which is typically used to avoid low-frequency noise~\cite{Yoshikawa2016Invited, Larsen2019Deterministic,Asavanant2019Generation}.
As shown in Figs.~\ref{fig:Fig4}c and d, these TWs are realized with comparable precision (temporal mode matching $\geq$92\%), without degrading the state quality as shown in Figs.~\ref{fig:Fig4}g and h.
Taken together, these results confirm that the TW is highly engineerable within a \SIrange{20}{100}{\nano\second} range.

As the final step,
we validate that the output quantum state is tunable independently of the TW.
In addition to the single-photon states already presented,
we generate two representative states for the four types of TWs introduced so far:
Schr\"odinger cat states as phase-sensitive states and two-photon states as higher-stellar-rank non-Gaussian states.
Their TWs consistently agree well with the targets regardless of the state type (Figs.~\ref{fig:Fig4}a--d), and all the Wigner functions exhibit negative regions with comparable quality (Figs.~\ref{fig:Fig4}i--l and m--p), thereby demonstrating the independent programmability of both the TW and the quantum state.

Across all the tomography results in Fig.~\ref{fig:Fig4}, the state qualities can be explained by $\sim$26\% loss within our experimental setup including imperfect measurement efficiency, and an additional loss arising from imperfect temporal-mode matching (refer to the legends in Figs.~\ref{fig:Fig4}a--d).
As a common trend, the TW mismatch primarily arises from a residual tail (clearly visible in Figs.~\ref{fig:Fig4}b and c, for example), which is attributed to the electrical high-pass filters in our measurement system (cutoff frequency: $\sim$\SI{159}{\kilo\hertz}).
Interestingly, this effect is relatively small for the DC-free balanced time-bin waveform in Fig.~\ref{fig:Fig4}d, resulting in the highest-quality states among four TWs for all three state types (Figs.~\ref{fig:Fig4}h, l, and p).
This indicates that our TW controllability is useful for optimizing the TW according to the system characteristics, thereby enabling more efficient and precise photonic quantum information processing.

\section{Discussion}

In conclusion, we have proposed a programmable non-Gaussian quantum light source with arbitrary and independent tunability of both the quantum state and the TW.
We have built a prototype and demonstrated the programmable generation of various non-Gaussian states with multiple TWs without any degradation in state quality.
This versatile light source is expected to offer diverse functionalities through its programmability, thereby accelerating the development of photonic quantum technologies.
Although the present demonstration was conducted on a tens-of-nanoseconds timescale, its operational range can, in principle, be extended.
Longer TWs can be realized by increasing the cavity time constant in the trigger path, whereas sub-nanosecond control is achievable via a modulation system exceeding \SI{1}{\giga\hertz}, potentially boosting the currently limited generation rate of non-Gaussian states by several orders of magnitude~\cite{Kawasaki2024Broadband}
Our approach generates target states synchronized with user-defined modulation timing, thereby enabling straightforward interfacing with other quantum devices, in contrast to previous CW-laser-based non-Gaussian light sources.
For example, when combined with loop-based quantum memories~\cite{Kaneda2015Time,Takeda2019demand} and appropriately synchronized, our light source could be upgraded to a semi-deterministic non-Gaussian quantum light source.

\section{Methods}
Figure~\ref{fig:Fig3} shows our experimental setup.
Here, two squeezed states at \SI{1545}{\nano\meter} are generated by pumping two broadband waveguide OPAs~\cite{Kashiwazaki2021Fabrication} with power-tunable CW light.
These states are then interfered on a BS with adjustable transmissivity.
Finally, the trigger light is sent to up to two superconducting nanostrip single-photon detectors (SNSPDs)~\cite{Miki2017Stable}.
This setup allows us to program the output state by tuning the two squeezing levels, the BS transmissivity, and the number of detected photons (one or two).
In parallel, an electro-optic intensity modulator (intensity EOM) and optical filters (a main filtering cavity together with two auxiliary higher-bandwidth filters) are introduced in the trigger path to realize our TW-shaping scheme.
This EOM is driven by a programmable electrical signal generated by an arbitrary waveform generator (AWG).
In our setup, the aforementioned requirement for the TW-shaping scheme is satisfied, as both the photon-pair correlation time ($\sim$\SI{20}{\pico\second}~\cite{Kawasaki2025Real}) and the SNSPD's temporal resolution (below \SI{100}{\pico\second}~\cite{Miki2017Stable}) are substantially shorter than the time constant of the main filtering cavity ($1/\gamma \approx \SI{50}{\nano\second}$).
The practically achievable TW timescale in our system is on the order of tens to hundreds of nanoseconds, determined by the bandwidth of our modulation system (\SI{330}{\mega\hertz}, limited by the AWG) and the cavity time constant.
The heralded states are evaluated only when the photon-detection times $T_1$ at SNSPD-1 and $T_2$ at SNSPD-2 satisfy $T_1 \geq T_{m2}$ for single-photon detection and $T_1, T_2 \geq T_{m2}$ for two-photon detection.
Since the state generation timing is synchronized with the modulation (not with the detection timing), the quadrature amplitude of the heralded state is measured using the TTL signal from the AWG as the trigger (Fig.~\ref{fig:Fig3}).

Note that our system reduces to a conventional one without TW control (corresponding to Fig.~\ref{fig:Fig2}a) by always keeping the EOM transmissivity at its maximum.
In this case, for single-photon detection, all SNSPD-1 detections are considered successful state-heralding events without applying the aforementioned success time window.
For two-photon detection, we define that state heralding is successful only if the detection time difference $|T_1 - T_2|$ is less than \SI{5}{\nano\second}.
These successful detections herald a state with an exponentially rising TW synchronized with the detection timing, as indicated in Eq.~\eqref{eq:f_inConventional}, and thus its quadrature is measured with the detection signal as the trigger.

\section{Acknowledgments}
We would like to thank Keitaro Anai for his insightful feedback on this manuscript.
This work was partly supported by the Japan Science and Technology Agency (JST) Grants No.~JPMJCR25I4, No.~JPMJFR223R, No.~JPMJMS2064, No.~JPMJMS256I, and No.~JPMJPF2221, and the Japan Society for the Promotion of Science (JSPS) KAKENHI Grants No.~23H01102, No.~23K17300, No.~24KJ0726, and No.25KJ0782. H.Tomoda and A.M. acknowledge financial support from The Forefront Physics and Mathematics Program to Drive Transformation (FoPM), WINGS Program, the University of Tokyo.

\section{Author contributions}
S.T. conceived the original concept and model framework, and supervised the project.
H.Tomoda formulated the detailed theoretical model with significant support from Y.N. and S.T.
H.Tomoda and Y.N. constructed the optical setup, performed the experiments, and analyzed the data with the assistance from A.M., D.O., and S.T.
T.K. and T.U. developed and provided the OPA modules.
S.M., M.Y., and H.Terai developed and provided the SNSPD system.
H.Tomoda wrote the manuscript under the guidance of S.T. with input from all authors.

\section{Competing interests}
The authors declare no competing interests.

\bibliography{MainText}

\end{document}



\title{Supplementary Information for\\
\textit{Programmable non-Gaussian quantum light source\\with state and temporal-waveform tunability}}

\maketitle
\tableofcontents


\section{Formulation of quantum state generation in proposed setup}
In this section, we formulate the generation of quantum states in the proposed setup.
We analyze the temporal waveform (TW) and the class of heralded non-Gaussian states in Sec.~\ref{sec:Heralding non-Gaussian states in the target TW} and Sec.~\ref{sec:Class of the heralded quantum states}, respectively.

\begin{figure*}[tbp]
  \centering
  \includegraphics[trim={9pt 0 78pt 0},clip, width=\linewidth]{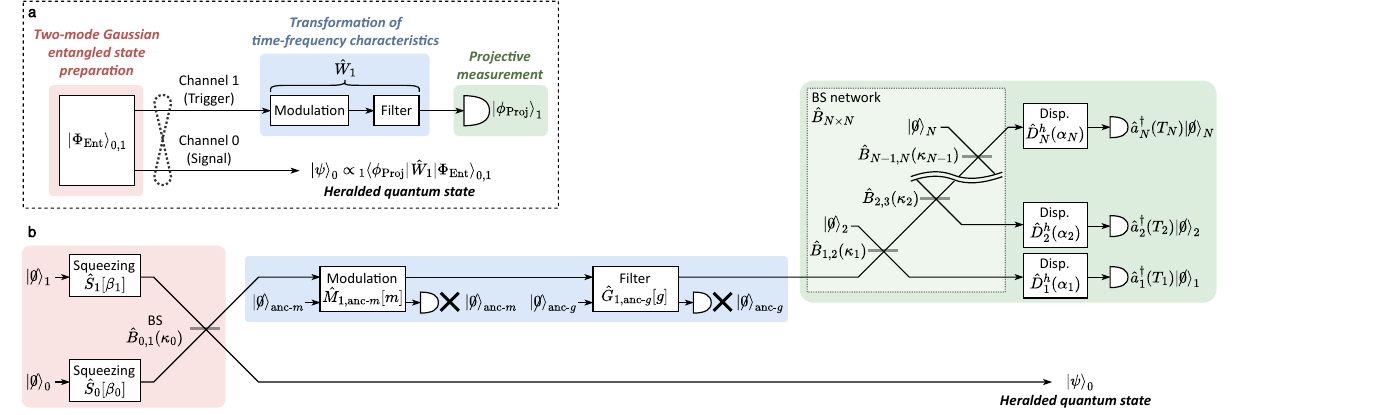}
  \caption{
    Programmable non-Gaussian quantum light source.
    a.~Schematic of the proposed scheme.
    b.~Detailed circuit diagram including all quantum operations.
    Disp. denotes displacement operations.
    \label{fig:S1}
    }
\end{figure*}

\subsection{Principle of temporal-waveform control\label{sec:Heralding non-Gaussian states in the target TW}}
Below, we show that the proposed quantum light source can herald a non-Gaussian state in a wave-packet mode defined by the target TW.
Figure~\ref{fig:S1} illustrates the setup with explicit representations of the quantum operations.
As shown in Fig.~\ref{fig:S1}a, the system consists of three stages: preparation of an entangled state $\ket{\Phi_\text{Ent}}_{0,1}$ in Channels~0 and 1; transformation $\hat{W}_{1}$ of the time-frequency characteristics in Channel~1 (a non-unitary operation occurring probabilistically under specific conditions); and a projective measurement $\ket{\phi_{\text{Proj}}}_1$ on Channel~1.
Based on these steps, the heralded state $\ket{\psi}_0$ in Channel~0 is given by
\begin{align}
  \ket{\psi}_0
  \propto
  \prescript{}{1}{\bra{\phi_{\text{Proj}}}}
  \hat{W}_{1}
  \ket{\Phi_{\text{Ent}}}_{0,1}.
  \label{eq:finally heralded state}
\end{align}
In the following, the target TW is denoted by $f_1(t)$.
We first examine how $\prescript{}{1}{\bra{\phi_{\text{Proj}}}} \hat{W}_{1}$ acts as a projector onto the wave-packet mode defined by $f_1$ in Secs.~\ref{sec:Projection}--\ref{sec:Projector}.
We then describe the preparation of the entangled state $\ket{\Phi_{\text{Ent}}}_{0,1}$, with an emphasis on the state purity in the mode $f_1$ in Sec.~\ref{sec:TwomodeGaussian}.
Finally, we show that a non-Gaussian state is heralded in this mode in Sec.~\ref{sec:FinalHeraldedStates}.

We adopt the following notation for wave-packet modes.
We first define the photon creation and annihilation operators associated with the wave-packet mode defined by the TW $f$ in Channel~$j$ as
\begin{align}
  \cre{j}[f] \coloneqq \int dt f(t) \cre{j} (t),
  \quad
  \ann{j}[f] \coloneqq \int dt f^{*}(t) \ann{j} (t), 
\end{align}
where $\cre{j}(t)$ denotes the instantaneous photon creation operator at time $t$ in Channel $j$.
The function $f(t)$ is normalized as $\int dt |f(t)|^2 = 1$, and the commutation relation $[\ann{j}(t), \cre{j}(t')] = \delta(t - t')$ implies $[\ann{j}[f], \cre{j}[f]] = 1$.
We then denote the quantum state $\ket{\psi}$ in the wave-packet mode of the TW $f$ in Channel~$j$ by $\ket{\psi}_j ^ {f}$.
For example, $\ket{n}_1^{f}$ represents the $n$-photon state in the mode $f$ in Channel~1.
In this section, all operators without a superscript (except for the photon creation and annihilation operators) act on multimode fields defined in the time-frequency domain.

\subsubsection{Projective measurement
\label{sec:Projection}}
Here we calculate the projective measurement $\prescript{}{1}{\bra{\phi_{\text{Proj}}}}$ in Channel~1 in Eq.~\eqref{eq:finally heralded state}.
To detect a total of $N$ photons using on/off detectors arranged in parallel, the beam in Channel~1 is split into $N$ channels (Channels~1--$N$) using beam splitters (BSs), as illustrated in the green region of Fig.~\ref{fig:S1}.
This process is modeled by successive BS operations incorporating Channels~2--$N$, which are initially the vacuum states.
The BS operation between adjacent Channels $j$ and $j+1$ ($j = 0, 1, \ldots, N-1$) is time-independent and defined by a real parameter $\kappa_j$ as
\begin{gather}
  \hat{B}_{j,j+1}(\kappa_j)
  \coloneqq
  \exp \left\{
  \frac{\kappa_j}{2}
  \int dt
  \left[
    \cre{j} (t) \ann{j+1} (t)
    - \cre{j+1} (t) \ann{j}(t)
    \right]
  \right\}.
  \label{eq:label3}
\end{gather}
Note that all integrals in this document, including that in Eq.~\eqref{eq:label3}, are taken over $(-\infty, \infty)$.
The BS network $\hat{B}_{N \times N}$ in Fig.~\ref{fig:S1} is then expressed as
\begin{gather}
  \hat{B}_{N \times N} \coloneqq
   \hat{B}_{N-1,N}(\kappa_{N-1}) \cdots \hat{B}_{2,3}(\kappa_2) \hat{B}_{1,2}(\kappa_1),
\end{gather}
and it transforms $\{\ann{j}(t)\}_{j=1} ^{N}$ as 
\begin{gather}
  \hat{B}_{N \times N} ^{\dagger} \ann{j}(t) \hat{B}_{N \times N}
  = \sum_{k=1} ^{N} c_{jk} \ann{k} (t),
  \label{eq:BS_1N_def}
\end{gather}
where we introduce the time-independent coefficients $\{c_{jk}\}_{j,k=1} ^{N}$.
Specifically, we consider the case where Channel~1 is equally split into $N$ channels by adjusting the parameters $\{\kappa_j\}_{j=1} ^{N-1}$.
In this case, the coefficients  $\{c_{j1}\}_{j=1} ^{N}$ satisfy $c_{j1}=1/\sqrt{N}$.

Following the BS network, a displacement is applied to a given wave-packet mode in each of Channels~1--$N$ by injecting modulated coherent light, as shown in Fig.~2 of the main text.
Such a displacement for the TW $h(t)$ (a normalized function) in Channel~$j \in \{1, 2,\dots, N\}$ is defined by the following operator:
\begin{align}
  \hat{D}_{j} ^{h} (\alpha_j)
  &
  \coloneqq
  \exp
  \left(
    \alpha_j \cre{j}[h] - \alpha_j ^{*} \ann{j}[h]
  \right),
\end{align}
where $\alpha_j \in \mathbb{C}$ denotes the displacement amplitude, and the specific profile of $h$ is given later in Eq.~\eqref{eq:cond3}.

Photon detection is then performed by an on/off detector at each channel.
We assume that the multiphoton components are negligible before the displacement in each channel, and that the displacement amplitudes are small, i.e., $|\alpha_j| \ll 1$, as in Ref.~\cite{Yukawa2013Generating}.
In this case, photon detection by the on/off detector can be considered single-photon detection.
Additionally by assuming that the detectors have infinite temporal resolution, photon detection at time $T_j$ in Channel~$j$ is described by a projection onto the state $\cre{j}(T_j)\vack{j}$, where $\vack{j}$ denotes the multimode vacuum state in Channel~$j$.
Accordingly, when photon detection occurs in all Channels~1--$N$, the projective measurement $\prescript{}{1}{\bra{\phi_{\text{Proj}}}}$ is expressed as
\begin{gather}
  \prescript{}{1}{\bra{\phi_{\text{Proj}}}}
  \propto
  \left(
    \bigotimes_{j=1} ^{N}
    \vacb{j}
    \ann{j} (T_j)
    \hat{D}_{j} ^{h} (\alpha_j)
    \right)
    \hat{B}_{N\times N}
    \left(
      \bigotimes_{j'=2} ^{N}
      \vack{j'}
      \right). \label{eq:projector}
\end{gather}
Under $|\alpha_j| \ll 1$, the term $\vacb{j} \ann{j}(T_j) \hat{D}_{j} ^{h} (\alpha_j)$ in Eq.~\eqref{eq:projector} is evaluated as follows:
\begin{align}
  \vacb{j}
  \ann{j}(T_j)
  \hat{D}_{j} ^{h} (\alpha_j)
  &
  \approx
  \vacb{j}
  \ann{j}(T_j)
  \left(
    1 + \alpha_j \cre{j}[h] - \alpha_j ^{*} \ann{j}[h]
  \right)
  \label{eq:DispExp_1}
  \\
  &
  \rightarrow 
  \vacb{j}
  \ann{j}(T_j)
  \left(
    1 + \alpha_j \int dt h(t) \cre{j}(t)
  \right)
  \label{eq:DispExp_2}
  \\
  &
  =
  \vacb{j}
  \left\{
    \ann{j}(T_j) 
      + \alpha_j \int dt h(t) 
    \left[
      \cre{j}(t) \ann{j}(T_j) + \delta (T_j - t)
    \right]
  \right\}
  \label{eq:DispExp_3}
  \\
  &
  =
  \vacb{j}
  \left[
    \ann{j}(T_j) + \alpha_j h (T_j)
  \right]
  .
  \label{eq:DispExp_4}
\end{align}
From Eq.~\eqref{eq:DispExp_1} to Eq.~\eqref{eq:DispExp_2}, the final term is neglected, as it corresponds to the detection of a two-photon component prior to displacement as a single photon, which is negligible under our assumption.
From Eq.~\eqref{eq:DispExp_2} to Eq.~\eqref{eq:DispExp_3}, the commutation relation $[\ann{j}(T_j), \cre{j}(t)] = \delta(T_j - t)$ is applied.
From Eq.~\eqref{eq:DispExp_3} to Eq.~\eqref{eq:DispExp_4}, $\vacb{j} \cre{j}(t) = 0$ is used.

By substituting Eqs.~\eqref{eq:BS_1N_def} and \eqref{eq:DispExp_4} into Eq.~\eqref{eq:projector}, the projector $\prescript{}{1}{\bra{\phi_{\text{Proj}}}}$ is finally obtained as
\begin{align}
  \prescript{}{1}{\bra{\phi_{\text{Proj}}}}
  &
  \approx
  \left\{
    \bigotimes_{j=1} ^{N}
    \vacb{j}
    \left[
      \ann{j}(T_j) + \alpha_j h (T_j)
    \right]
  \right\}
  \hat{B}_{N \times N}
  \left(
    \bigotimes_{j'=2} ^{N}
    \vack{j'}
  \right)
  \label{eq:proj_2}
  \\
  &
  =
  \bigotimes_{j=1} ^{N}
  \vacb{j}
  \left\{
    \bigotimes_{j''=1} ^{N}
    \hat{B}_{N \times N} ^{\dagger}
    \left[
      \ann{j''}(T_{j''})
      + \alpha_{j''} h (T_{j''})
    \right]
    \hat{B}_{N \times N}
  \right\}
  \bigotimes_{j'=2} ^{N}
  \vack{j'}
  \label{eq:proj_3}
  \\
  & 
  =
  \bigotimes_{j=1} ^{N}
  \vacb{j}
  \left\{
  \bigotimes_{j''=1} ^{N}
  \left[
  \left(
    \sum_{k=1} ^{N} c_{j''k} \ann{k} (T_{j''})
  \right)
    + \alpha_{j''} h (T_{j''})
  \right]
  \right\}
  \bigotimes_{j'=2} ^{N}
  \vack{j'}
  \label{eq:proj_41}
  \\
  & 
  =
  \bigotimes_{j=1} ^{N}
  \vacb{j}
  \left[
  \left( 
  \frac{1}{\sqrt{N}}
  \ann{1} (T_1) + 
    \sum_{k=2} ^{N} c_{1k} \ann{k} (T_1)
  \right)
    + \alpha_1 h (T_1)
  \right]
  \nonumber \\
  & \hspace{20mm}
  \left[
  \left( 
  \frac{1}{\sqrt{N}}
  \ann{1} (T_2) + 
    \sum_{k=2} ^{N} c_{2k} \ann{k} (T_2)
  \right)
    + \alpha_2 h (T_2)
  \right]
  \nonumber \\
  & \hspace {30mm}
  \cdots
  \left[
    \left( 
  \frac{1}{\sqrt{N}}
    \ann{1} (T_N) + 
    \sum_{k=2} ^{N} c_{Nk} \ann{k} (T_N)
  \right)
    + \alpha_N h (T_N)
  \right]
  \bigotimes_{j'=2} ^{N}
  \vack{j'}
  \label{eq:proj_4}
  \\
  & 
  =
  \vacb{1}
  \bigotimes_{j=1} ^{N}
  \left[
    \frac{1}{\sqrt{N}} 
    \ann{1} (T_{j})
    + \alpha_{j} h (T_{j})
  \right].
  \label{eq:proj_6}
\end{align}
From Eq.~\eqref{eq:proj_2} to Eq.~\eqref{eq:proj_3}, we used $\left(\bigotimes_{j=1} ^{N} \vacb{j}\right)\hat{B}_{N \times N} ^{\dagger} = \bigotimes_{j=1} ^{N} \vacb{j}$ together with $\hat{B}_{N \times N} \hat{B}_{N \times N} ^{\dagger} = \hat{I}$ (the identity operator).
From Eq.~\eqref{eq:proj_4} to Eq.~\eqref{eq:proj_6}, the terms containing $\{\ann{k} (T_{1})\}_{k=2}^{N}$, $\{\ann{k} (T_{2})\}_{k=2}^{N}$, $\dots$, $\{\ann{k} (T_{N})\}_{k=2}^{N}$ vanish because they are sandwiched between vacuum states in Channels~2--$N$.

\subsubsection{Transformation of time-frequency characteristics
\label{Sec:transformation}}

Next, we consider the transformation $\hat{W}_1$ of time-frequency characteristics, realized through modulation and filtering as shown in Fig.~\ref{fig:S1}.
Following Ref.~\cite{Takase2022Quantum}, transmission through a filter in Channel~1 can be formulated as a quantum operation involving two channels:
\begin{align}
  \hat{G}_{1,\text{anc-}g} ^{\dagger} [g] \ann{1}(t) \hat{G}_{1,\text{anc-}g} [g] 
  = 
  \int d\tau \left\{
    g(\tau) \ann{1} (t-\tau) -
  \left[
    \delta(\tau) - g(\tau) 
    \right]
    \ann{\text{anc-}g}(t-\tau)
    \right\}
    ,
\end{align}
where Channel~$\text{anc-}g$ refers to an ancillary channel initialized in the vacuum state.
The function $g(t)$ represents the normalized impulse response of the filter.
Similarly, the intensity modulation operation for Channel~$j$, involving Channel~$\text{anc-}m$, an ancillary channel initialized in the vacuum state, is defined as
\begin{align}
  \hat{M}_{1,\text{anc-}m} ^{\dagger} [m] \ann{1}(t) \hat{M}_{1,\text{anc-}m} [m]
  =
  \sin{[m(t)]} \ann{1} (t) + \cos{[m(t)]} \ann{\text{anc-}m} (t).
\end{align}
This can be interpreted as a BS with a time-varying transmissivity $|\sin{[m(t)]}|^2$, and we hereafter refer to $\sin{[m(t)]}$ as the modulation function.

Consequently, the quantum operation $\hat{W}_1$ is written as
\begin{align}
  \hat{W}_1
  \propto
  \vacb{\text{anc-}m}
  \vacb{\text{anc-}g}
  \hat{G}_{1,\text{anc-}g} [g]
  \hat{M}_{1,\text{anc-}m} [m]
  \vack{\text{anc-}g}
  \vack{\text{anc-}m},
  \label{eq:projection}
\end{align}
where we assume that two ancillary channels ($\text{anc-}g$ and $\text{anc-}m$) are projected onto vacuum states.
This projection can be omitted experimentally when the vacuum component in Channel~1 is dominant prior to this operation.

To calculate $\prescript{}{1}{\bra{\phi_{\text{Proj}}}} \hat{W}_{1}$ in the next section, we derive the combined effect of the above two operations, $\hat{G}_{1,\text{anc-}g}[g]$ and $\hat{M}_{1,\text{anc-}m}[m]$, on $\ann{1}(t)$ as follows:
\begin{align}
  &
  \hat{M}_{1,\text{anc-}m} ^{\dagger} [m]
  \hat{G}_{1,\text{anc-}g} ^{\dagger} [g]
  \ann{1} (t)
  \hat{G}_{1,\text{anc-}g} [g]
  \hat{M}_{1,\text{anc-}m} [m]
  \nonumber
  \\
  &=
  \hat{M}_{1,\text{anc-}m} ^{\dagger} [m]
  \left(
  \int d\tau
  \left\{
    g(\tau) \ann{1} (t-\tau) -
  \left[
    \delta(\tau) - g(\tau) 
    \right]
    \ann{\text{anc-}g}(t-\tau)
    \right\}
  \right)
  \hat{M}_{1,\text{anc-}m} [m]
  \label{eq:TMtransformation_2}
  \\
  &=
  \int d\tau
  \left\{
    g(\tau) \hat{M}_{1,\text{anc-}m} ^{\dagger} [m] \ann{1} (t-\tau)  \hat{M}_{1,\text{anc-}m} [m]
    -
  \left[
    \delta(\tau) - g(\tau) 
    \right]
    \ann{\text{anc-}g}(t-\tau)
    \right\}
    \label{eq:TMtransformation_3}
  \\
  &=
  \int d\tau
  \Bigl(
    g(\tau) 
    \bigl\{
    \sin{[m(t-\tau)]} \ann{1} (t-\tau) + \cos{[m(t-\tau)]} \ann{\text{anc-}m} (t-\tau)
    \bigr\}
    -
  \left[
    \delta(\tau) - g(\tau) 
    \right]
    \ann{\text{anc-}g}(t-\tau)
  \Bigr)
  \label{eq:TMtransformation_4}
  \\
  &=
  \int d t'
  \Bigl(
    g(t-t') 
    \bigl\{
    \sin{[m(t')]} \ann{1} (t') + \cos{[m(t')]} \ann{\text{anc-}m} (t')
    \bigr\}
    -
  \left[
    \delta(t-t') - g(t-t') 
    \right]
    \ann{\text{anc-}g}(t')
  \Bigr).
  \label{eq:TMtransformation_5}
\end{align}
From Eq.~\eqref{eq:TMtransformation_4} to Eq.~\eqref{eq:TMtransformation_5}, we changed the variable from $\tau$ to $t'$ as $\tau = t - t'$.

\subsubsection{Projector onto quantum state in target temporal waveform
\label{sec:Projector}}
We then calculate the projector incorporating the transformation of the time-frequency characteristics, denoted as $\prescript{}{1}{\bra{\phi_{\text{Proj}}}} \hat{W}_{1}$.
Based on the results in Secs.~\ref{sec:Projection} and \ref{Sec:transformation}, this is given as follows:
\begin{align}
  &
  \prescript{}{1}{\bra{\phi_{\text{Proj}}}}
  \hat{W}_{1}
  \nonumber
  \\
  & \propto
  \vacb{1}
  \left\{
  \bigotimes_{j=1} ^{N}
  \left[
    \frac{1}{\sqrt{N}}   
    \ann{1} (T_j)
    + \alpha_j h (T_j)
  \right]
  \right\}
  \vacb{\text{anc-}m}
  \vacb{\text{anc-}g}
  \hat{G}_{1,\text{anc-}g} [g]
  \hat{M}_{1,\text{anc-}m} [m]
  \vack{\text{anc-}g}
  \vack{\text{anc-}m}
  \label{eq:phiW_1}
  \\
  & =
  \vacb{1}
  \vacb{\text{anc-}m}
  \vacb{\text{anc-}g}
  \left\{
  \bigotimes_{j=1} ^{N}
  \hat{M}_{1,\text{anc-}m} ^{\dagger} [m]
  \hat{G}_{1,\text{anc-}g} ^{\dagger} [g]
  \left[
     \frac{1}{\sqrt{N}}
     \ann{1} (T_j) 
    + \alpha_j h (T_j)
  \right]
  \hat{G}_{1,\text{anc-}g} [g]
  \hat{M}_{1,\text{anc-}m} [m]
  \right\}
  \vack{\text{anc-}g}
  \vack{\text{anc-}m}
  \label{eq:phiW_2}
  \\
  & =
  \vacb{1}
  \vacb{\text{anc-}m}
  \vacb{\text{anc-}g}
  \left\{
  \bigotimes_{j=1} ^{N}
  \left[
    \frac{1}{\sqrt{N}}
    \hat{M}_{1,\text{anc-}m} ^{\dagger} [m]
    \hat{G}_{1,\text{anc-}g} ^{\dagger} [g]
    \ann{1} (T_j) 
    \hat{G}_{1,\text{anc-}g} [g]
    \hat{M}_{1,\text{anc-}m} [m]
    + \alpha_j h (T_j)
  \right]
  \right\}
  \vack{\text{anc-}g}
  \vack{\text{anc-}m}
  \label{eq:phiW_3}
  \\
  & =
  \vacb{1}
  \vacb{\text{anc-}m}
  \vacb{\text{anc-}g}
  \left\{
  \bigotimes_{j=1} ^{N}
  \left[
    \frac{1}{\sqrt{N}}
  \int d t'
  \Bigl(
    g(T_j-t') 
    \bigl\{
    \sin{[m(t')]} \ann{1} (t') + \cos{[m(t')]} \ann{\text{anc-}m} (t')
    \bigr\}
  \right.
  \right.
  \nonumber \\*
  & \hspace{65mm}
  -
  \left[
    \delta(T_j-t') - g(T_j-t') 
    \right]
    \ann{\text{anc-}g}(t')
  \Bigr)
    + \alpha_j h (T_j)
  \biggr]
  \Biggr\}
  \vack{\text{anc-}g}
  \vack{\text{anc-}m}
  \label{eq:phiW_4}
  \\
  & =
  \vacb{1}
  \bigotimes_{j=1} ^{N}
  \left\{
    \left[
      \frac{1}{\sqrt{N}}
      \int d t
      g(T_j-t) 
      \sin{[m(t)]} \ann{1} (t)
      \right]
    + \alpha_j h (T_j)
  \right\}
  .
  \label{eq:phiW_5}
\end{align}
From Eq.~\eqref{eq:phiW_1} to Eq.~\eqref{eq:phiW_2}, we used the fact that  $\hat{M}_{1,\text{anc-}m} ^{\dagger} [m] \hat{G}_{1,\text{anc-}g} ^{\dagger} [g]$ leaves the state $\vacb{1} \vacb{\text{anc-}m} \vacb{\text{anc-}g}$ unchanged.
Additionally, we inserted the identity operator $\hat{G}_{1,\text{anc-}g} [g] \hat{M}_{1,\text{anc-}m} [m] \hat{M}_{1,\text{anc-}m} ^{\dagger} [m] \hat{G}_{1,\text{anc-}g} ^{\dagger} [g]$ between the elements of the tensor product to apply the result in Eq.~\eqref{eq:TMtransformation_5}.
Finally, from Eq.~\eqref{eq:phiW_4} to Eq.~\eqref{eq:phiW_5}, the terms containing the operators $\ann{\text{anc-}m}(t')$ and $\ann{\text{anc-}g}(t')$ vanish because they are sandwiched between vacuum states in these channels.

\begin{figure}[!b]
  \centering
  \includegraphics[trim={20mm 0 0 0},clip, width=0.32\linewidth]{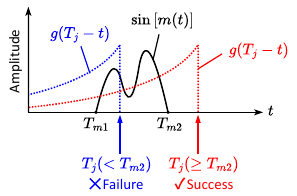}
  \caption{
  Temporal relationship between the modulation function $\sin{[m(t)]}$ and the photon-detection times $\{T_j\}_{j=1} ^{N}$.
  \label{fig:S2}
  }
\end{figure}

In general, the expression for the projector $\prescript{}{1}{\bra{\phi_{\text{Proj}}}}\hat{W}_{1}$ in Eq.~\eqref{eq:phiW_5} indicates that which modes are projected onto which states depends on the photon-detection times $\{T_j\}_{j=1} ^{N}$.
This dependence can, however, be removed by imposing several assumptions.
First, we assume that the modulation function $\sin{[m(t)]}$ is defined in the interval $T_{m1} \le t \le T_{m2}$, and zero elsewhere, as shown in Fig.~\ref{fig:S2}.
We then impose the following three requirements:
\begin{align}
  &
  T_j \geq T_{m2} \quad (\text{for } j=1,2,\dots, N)
  ,
  \label{eq:cond1}
  \\
  &
  g(t) \propto
  \begin{dcases}
    e^{-\gamma t} & (t \geq 0), \\    
    0 & (t < 0), \\ 
  \end{dcases}
  \label{eq:cond2}
  \\
  &
  h(t) = g(t-T_{m2})
  \label{eq:cond3}.
\end{align}
In this case, Eq.~\eqref{eq:phiW_5} can be rearranged to
\begin{align}
  \prescript{}{1}{\bra{\phi_{\text{Proj}}}}
  \hat{W}_{1}
  &
  \propto
  \vacb{1} 
  \bigotimes_{j=1} ^{N}
  \left\{
    \left[
    \frac{e^{-\gamma T_j}}{\sqrt{N}}
  \int d t
  e^{\gamma t}
    \sin{[m(t)]} \ann{1} (t)
    \right]
  + \alpha_j
  e^{-\gamma (T_j-T_{m2})}
  \right\}
  \label{eq:rewritephiW_pre}
  \\
  &
  \propto
  \vacb{1} 
  \bigotimes_{j=1} ^{N}
  \left\{
  \left[
    \frac{1}{\sqrt{N}}
  \int d t
    e^{\gamma (t-T_{m2})} 
    \sin{[m(t)]} \ann{1} (t)
  \right]
    + \alpha_j
  \right\}
  ,
  \label{eq:rewritephiW}
\end{align}
which is independent of $\{T_j\}_{j=1} ^{N}$ up to a multiplicative constant.
This independence can be understood as the result of the following reasons. 
By applying the conditions in Eqs.~\eqref{eq:cond1} and \eqref{eq:cond2}, the terms containing $T_j$ can be factorized as $e^{-\gamma T_j}$ from the integral
$\int dt g(T_j-t)  \sin{[m(t)]} \ann{1} (t)$, as shown in Eq.~\eqref{eq:rewritephiW_pre}. 
This eliminates the $\{T_j\}_{j=1} ^{N}$-dependence of the TW. 
At the same time, by performing the displacement for the cavity response profile $g$ as specified in Eq.~\eqref{eq:cond3}, the ratio between the two terms inside the curly brackets of Eq.~\eqref{eq:rewritephiW_pre} becomes fixed. 
As a result, the projector maps states onto the same superposition regardless of $\{T_j\}_{j=1} ^{N}$. 
Consequently, the $\{T_j\}_{j=1}^{N}$-dependence is eliminated from $\prescript{}{1}{\bra{\phi_{\text{Proj}}}}\hat{W}_{1}$ in Eq.~\eqref{eq:rewritephiW}, and the projected mode and state do not vary with $\{T_j\}_{j=1}^{N}$.

To realize a projective measurement for the target TW $f_1$ (real-valued function) under the above conditions, we tune the modulation function $\sin{[m(t)]}$ such that the target TW $f_1$ can be expressed as
\begin{align}
  f_1(t) 
  =
  c'
  e^{\gamma (t-T_{m2})} 
  \sin{[m(t)]}
  \label{eq:relationfandm}
  ,
\end{align}
where $c'$ is a normalization factor.
In this case, we can define an orthonormal set of functions $\{f_l\}_{l=1} ^{\infty}$ that includes $f_1$, in order to decompose the multimode vacuum state as $\vacb{1}=\otimes_{l=1}^{\infty} \prescript{f_l}{1}{\bra{0}}$.
Then, Eq.~\eqref{eq:rewritephiW} is rewritten as 
\begin{align}
  \prescript{}{1}{\bra{\phi_{\text{Proj}}}}
  \hat{W}_{1}
  &
  \propto
  \vacb{1} 
  \bigotimes_{j=1} ^{N}
  \left(
    \frac{1}{\sqrt{N}c'}
    \ann{1} [f_1]
    + \alpha_j
  \right)
  \label{eq:finalprojector_pre}
  \\
  & \propto
  \left(
    \sum_{n=0} ^{N}
    C'_n(\alpha_1, \alpha_2, \dots, \alpha_N)
    \prescript{f_1}{1}{\bra{n}}
  \right)
  \left(
    \bigotimes_{l=2} ^{\infty}
    \prescript{f_l}{1}{\bra{0}}
  \right),
  \label{eq:finalprojector}
\end{align}
by introducing the coefficients $\{C'_n(\alpha_1, \alpha_2, \dots, \alpha_N)\}_{n=1} ^{N}$.
Equation~\eqref{eq:finalprojector} shows that $\prescript{}{1}{\bra{\phi_{\text{Proj}}}} \hat{W}_{1}$ projects the state in the mode $f_1$ onto a superposition of Fock states up to $N$ photons.
As a result, we can conclude that the projected mode and state can be independently engineered by
appropriately tuning $\sin{[m(t)]}$ according to Eq.~\eqref{eq:relationfandm} and by adjusting the displacement amplitudes $\{\alpha_j\}_{j=1}^{N}$, respectively.

\subsubsection{Two-mode Gaussian entangled state preparation
\label{sec:TwomodeGaussian}}
So far, we have shown that $\prescript{}{1}{\bra{\phi_{\text{Proj}}}} \hat{W}_{1}$ can be decomposed into a component acting on the mode $f_1$ and components acting on the remaining modes $\{f_l\}_{l=2} ^{\infty}$.
We now focus on the entangled states $\ket{\Phi_{\text{Ent}}}_{0,1}$ on which the projector acts, and discuss a similar decomposition into the mode $f_1$ and the remaining modes $\{f_l\}_{l=2} ^{\infty}$.

In the typical heralding scheme for non-Gaussian state generation, Gaussian entangled states, such as Einstein-Podolsky-Rosen states, are employed as the initial resource because they can be deterministically generated using a simple experimental setup.
In our light source (Fig.~\ref{fig:S1}), we prepare such an initial entangled resource by interfering two squeezed states at a variable-transmissivity BS.
Mathematically, this can be expressed as
\begin{align}
  \ket{\Phi_{\text{Ent}}}_{0,1}
  =
  \hat{B}_{0,1}(\kappa_0)
  \hat{S}_0[\beta_0]
  \hat{S}_1[\beta_1]
  \vack{0}
  \vack{1},
  \label{eq:two-mode entangled states}
\end{align}
where $\hat{B}_{0,1}(\kappa_0)$ is the BS operator introduced in Eq.~\eqref{eq:label3}, and $\hat{S}_j[\beta_j] \ (j=0,1)$ denotes the squeezing operator, which will be defined later.
These operators are multimode with respect to time-frequency properties, and below we express them using the orthonormal set of wave-packet modes $\{f_l\}_{l=1} ^{\infty}$.

First, following Ref.~\cite{Takase2022Quantum}, we decompose the BS operator into a direct product of BS operators acting on the orthonormal set of modes $\{f_l\}_{l=1} ^{\infty}$ as
\begin{align}
  \hat{B}_{0,1}(\kappa_0)
  & =
  \bigotimes_{l=1}^{\infty}
  \exp \left[
    \frac{\kappa_0}{2}
    \left(
    \cre{0}[f_l] \ann{1}[f_l]
    - \cre{1}[f_l] \ann{0}[f_l]
    \right)
    \right]
  \coloneqq
  \bigotimes_{l=1}^{\infty}
  \hat{B}_{0,1} ^{f_l} (\kappa_0).
  \label{eq:BS_4}
\end{align}
Second, the squeezing operator $\hat{S}_j[\beta_j]$ is defined in terms of the photon-pair creation and annihilation operators, $\hat{P}_j^{\dagger}[\beta_j]$ and $\hat{P}_j[\beta_j]$, as~\cite{Yoshikawa2017Purification}
\begin{gather}
  \hat{S}_j[\beta_j]
  \coloneqq 
  \exp
  \left[
    \frac{1}{2} 
    \left(
      \hat{P}_j  [\beta_j] 
      - \hat{P}^{\dagger} _j[\beta_j]
    \right)
  \right],
  \label{eq:Sqop}
  \\
  \hat{P}_j ^{\dagger} [\beta_j]
  \coloneqq
  \int dt \int dt' \beta_j (t-t') \cre{j} (t) \cre{j} (t'),
\end{gather}
where $\beta_j(t-t')$ denotes the temporal correlation function of the generated photon pairs.
We then focus on squeezing in the mode $f_1$ and examine its purity, following Ref.~\cite{Yoshikawa2017Purification}.
To this end, we expand $\hat{P}_j ^{\dagger} [\beta_j]$ in terms of the photon creation operators for the wave-packet modes $\{\cre{j} [f_l]\}_{l=1} ^{\infty}$ as
\begin{gather}
  \hat{P}_j {}^{\dagger} [\beta_j]
  =
  p_{11} \cre{j}[f_1] ^2 
  + 2 p_{12} \cre{j}[f_1] \cre{j}[f_2]
  + \hat{P}' _j {}^{\dagger} ,
  \label{eq:pjdecomposition}
  \\
  \begin{dcases}
    p_{11} \coloneqq
    \int dt f_1 ^{*} (t) (f_1 ^{*} * \beta_j )(t)
    ,
    \\
    p_{12} \coloneqq
    \sqrt{\|f_1 ^{*} * \beta_j\|^2 - |p_{11}|^2}
    .
  \end{dcases}
\end{gather}
Here, $f^{*} _1 * \beta_j$ denotes the convolution of the functions $f_1 ^{*}$ and $\beta_j$, $\|f_1 ^{*} * \beta_j\|$ denotes the norm of $f_1 ^{*} * \beta_j$, and $\hat{P}'_j{}^{\dagger}$ commutes with $\ann{j}[f_1]$, but not necessarily with $\{\ann{j}[f_l]\}_{l=2}^{\infty}$.
In Eq.~\eqref{eq:pjdecomposition}, the first term represents photon-pair creation within a single wave-packet mode (single-mode squeezing), whereas the second term represents photon-pair creation across two different wave-packet modes (two-mode squeezing).
To suppress the second term and realize pure squeezing for the TW $f_1$, we ensure $|p_{12}|\ll 1$ by choosing $\beta_j(t-t')$ as a sufficiently narrow function of $t-t'$ compared with $f_1$.
When $\beta_j(t-t')$ is approximated as $\beta_j(t-t')\approx r_j \delta(t-t')$ with a complex coefficient $r_j$, and $f_1$ is real-valued, we obtain $p_{11}=r_j$ and $p_{12}\approx 0$.
Under this condition, the multimode squeezing operator in Eq.~\eqref{eq:Sqop} is approximately decomposed into the single-mode pure squeezing operator  $\hat{S}_j ^{f_1} (r_j)$ for the TW $f_1$ and an operator $\hat{U}^{\{f_2, f_3, \dots\}}$ for the remaining TWs $\{f_l\}_{l=2} ^{\infty}$:
\begin{gather}
  \hat{S}_j[\beta_j] \approx \hat{S}_j ^{f_1} (r_j) \hat{U}_j^{\{f_2, f_3, \dots\}},
  \label{eq:final squeezing operator}
  \\
  \hat{S}_j ^{f_1} (r_j) \coloneqq 
  \exp
  \left\{
    \frac{1}{2} 
    \left[
      r_j ^{*} \left( \ann{j}[f_1] \right)^2 
      - r_j \left(\cre{j}[f_1]\right)^2
    \right]
  \right\}
  .
\end{gather}

Finally, combining Eqs.~\eqref{eq:BS_4} and \eqref{eq:final squeezing operator}, the two-mode entangled state in Eq.~\eqref{eq:two-mode entangled states} can be rewritten as
\begin{align}
  \ket{\Phi_{\text{Ent}}}_{0,1}
  & \approx
  \left(
    \bigotimes_{l=1}^{\infty}
    \hat{B}_{0,1} ^{f_l} (\kappa_0)
  \right)
  \hat{S}_0 ^{f_1} (r_0) \hat{U}_0 ^{\{f_2, f_3, \dots\}}
  \hat{S}_1 ^{f_1} (r_1) \hat{U}_1 ^{\{f_2, f_3, \dots\}}
  \vack{0}
  \vack{1}
  \\
  & =
  \hat{B}_{0,1} ^{f_1} (\kappa_0)
  \hat{S}_0 ^{f_1} (r_0)
  \hat{S}_1 ^{f_1} (r_1)
  \ket{0}_0 ^{f_1}
  \ket{0}_1 ^{f_1}
  \hat{U'}_{0,1}^{\{f_2, f_3, \dots\}}
  \left(
    \bigotimes_{l=2}^{\infty}
    \ket{0}_0 ^{f_l} 
    \ket{0}_1 ^{f_l} 
  \right),
  \label{eq:entangledstate}
\end{align}
where we absorb the operators acting on the TWs $\{f_l\}_{l=2} ^{\infty}$ into a new operator $\hat{U'}_{0,1}^{\{f_2, f_3, \dots\}}$.
Consequently, the entangled state $\ket{\Phi_{\text{Ent}}}_{0,1}$ can be decomposed into the state associated with the target TW $f_1$ and the state associated with the remaining TWs $\{f_l\}_{l=2} ^{\infty}$.
As described in the main text with reference to Fig.~2d, our scheme assumes the case of interfering two orthogonally squeezed states, so that we can restrict the squeezing parameters $r_0$ and $r_1$ to real values in the following discussion.

\subsubsection{Final heralded states
\label{sec:FinalHeraldedStates}}
In this section, we derive the final heralded state using the results of the projector in Eq.~\eqref{eq:finalprojector} and the two-mode entangled state in Eq.~\eqref{eq:entangledstate}.
The final state, $\ket{\psi}_0$ defined in Eq.~\eqref{eq:finally heralded state}, is expressed as follows:
\begin{align}
  \ket{\psi}_0
  & \propto
  \underbrace{
  \left(
    \sum_{n=0} ^{N}
    C'_n(\alpha_1, \alpha_2, \dots, \alpha_N)
    \prescript{f_1}{1}{\bra{n}}
  \right)
  \hat{B}_{0,1} ^{f_1} (\kappa_0)
  \hat{S}_0 ^{f_1} (r_0)
  \hat{S}_1 ^{f_1} (r_1)
  \ket{0}_0 ^{f_1}
  \ket{0}_1 ^{f_1}
  }
  _{\text{State in the target TW $f_1$}}
  \underbrace{
  \left(
    \bigotimes_{l=2} ^{\infty}
    \prescript{f_l}{1}{\bra{0}}
  \right)
  \hat{U'}_{0,1} ^{\{f_2, f_3, \dots\}}
  \left(
    \bigotimes_{l=2}^{\infty}
    \ket{0}_0 ^{f_l} 
    \ket{0}_1 ^{f_l} 
  \right)}
  _{\text{State in the remaining TWs $\{f_l\}_{l=2} ^{\infty}$}}
  \label{eq:final_heralded_state}
\end{align}
As indicated in Eq.~\eqref{eq:final_heralded_state}, the non-Gaussian state is heralded in the target TW $f_1$, whereas the Gaussian state in the remaining TWs $\{f_l\}_{l=2} ^{\infty}$.
This expression means that the type of the heralded state in the target TW $f_1$ depends on both the preparation of the two-mode Gaussian entangled state and the projective measurement.

In summary, we have shown that our light source can herald a non-Gaussian state in the target TW $f_1$ by properly tuning the modulation function $\sin{[m(t)]}$.  
The type of the non-Gaussian state can also be independently controlled by adjusting both the parameters $(r_0, r_1, \kappa_0)$ in the entangled state preparation and the set of displacement amplitudes $\{\alpha_j\}_{j=1}^{N}$ in the projective measurement.

As a remark, we discuss the fundamental limits on the timescale of the TW.
Our discussion so far relies on the following two assumptions:
\begin{enumerate}
\item The temporal correlation function $\beta_j(t-t')\ (j=0,1)$ is sufficiently narrow compared to the target TW $f_1$, such that $\beta_j(t-t') \approx r_j \delta(t-t')$.
\item The detectors have effectively infinite temporal resolution.
This condition is approximately satisfied when the resolution is much shorter than the timescales of both the target TW $f_1$ and the filter response time constant $1/\gamma$.
\end{enumerate}
Under these assumptions, we can realize a TW whose timescale is sufficiently longer than both the photon-pair correlation time and the detector resolution.
Conversely, since the TW $f_1$ is given by the product of $e^{\gamma t}$ and $\sin[m(t)]$ in Eq.~\eqref{eq:relationfandm}, its duration is limited by the time constant $1/\gamma$.
In principle, a TW longer than $1/\gamma$ can be realized by appropriately tuning $\sin[m(t)]$; however, this is less practical because the success probability of TW shaping decreases significantly.

\subsection{Principle of state control\label{sec:Class of the heralded quantum states}}
In Sec.~\ref{sec:Heralding non-Gaussian states in the target TW}, we confirmed that non-Gaussian states are heralded in the desired TW in the proposed setup; however, the specific forms of these states have not yet been discussed.
Here, we show that arbitrary quantum states of the form $\hat{S}_0 (r_{\text{out}}) \sum_{n=0}^{N} C_n \ket{n}_0$ ($r_{\text{out}} \in \mathbb{R}$, $C_0, C_1, \dots, C_N \in \mathbb{C}$; the labels regarding the TW are omitted) can be generated in the case of $N$ photon detections.
By further applying experimentally feasible displacement and phase-shift operations to these states, arbitrary stellar-rank-$N$ states can be generated.

\subsubsection{Type of heralded state\label{sec:statetype}}
Here, we derive an expression for the final heralded state $\ket{\psi}_0$ in the target TW $f_1$ and show that it can always be written in the form $\hat{S}_0 (r_{\text{out}}) \sum_{n=0}^{N} C_n \ket{n}_0$.
From Eq.~\eqref{eq:final_heralded_state}, the final state $\ket{\psi}_0$ can be expressed as
\begin{gather}
  \ket{\psi}_0
  \propto
  \sum_{n=0} ^{N}
  C'_n
  \prescript{}{1}{\braket{n|\Phi_{\text{Ent}}}}_{0,1}
  ,
  \label{eq:psi0_1}
  \\
  \ket{\Phi_{\text{Ent}}}_{0,1}
  =  
  \hat{B}_{0,1} (\kappa_0)
  \hat{S}_0 (r_0)
  \hat{S}_1 (r_1)
  \ket{0}_0
  \ket{0}_1, 
  \label{eq:entangledstate_singlemode}
\end{gather}
where we omit the TW labels in state vectors and operators, as well as the parameters $\{\alpha_j\}_{j=1}^{N}$ in $\{C'_n\}_{n=0} ^N$.
The state $\prescript{}{1}{\braket{n|\Phi_{\text{Ent}}}}_{0,1}$ represents the state heralded by an $n$-photon detection; we define the corresponding normalized state  $\ket{\psi^{(n)}}_0$ as
\begin{align}
  \ket{\psi^{(n)}}_0
  =
  \frac{1}{\sqrt{P(n)}}
  \prescript{}{1}{\braket{n|\Phi_{\text{Ent}}}}_{0,1}
  ,
  \label{eq:n-photon_projected_normalized}
\end{align}
by introducing the $n$-photon detection probability $P(n)$ as
\begin{align}
  P(n) = 
  \prescript{}{0,1}{\braket{\Phi_{\text{Ent}}|n}}_{1}
  \prescript{}{1}{\braket{n|\Phi_{\text{Ent}}}}_{0,1}.
  \label{eq:eq:n-photon_probability}
\end{align}
Consequently, the state in Eq.~\eqref{eq:psi0_1} can be expressed as
\begin{align}
  \ket{\psi}_0
  & \propto
  \sum_{n=0} ^{N}
  C'_n
  \sqrt{P(n)}
  \ket{\psi^{(n)}}_0.
  \label{eq:superposition}
\end{align}
Here, we assume that $\ket{\Phi_{\text{Ent}}}_{0,1}$ is an entangled state.
Under this assumption, the state in Channel 1, obtained by tracing out Channel 0, is a squeezed thermal state (excluding the squeezed vacuum state). Such states contain all photon-number components, so that none of $\{P(n)\}_{n=0}^{N}$ vanishes.
We also note that $C'_N \neq 0$ when $N$ photons are detected, since $C'_N$ is introduced from Eq.~\eqref{eq:finalprojector_pre} to Eq.~\eqref{eq:finalprojector}.
Below, we first evaluate the states  $\ket{\psi^{(n)}}_0$ using wave functions and then discuss their superposition $\ket{\psi}_0$.

In our formulation, we define the quadrature amplitude $\hat{x}_j$ in Channel~$j$ as
\begin{align}
  \hat{x}_j = \frac{\ann{j} + \cre{j}}{\sqrt{2}} \quad (\hbar=1),
\end{align}
and derive the normalized wave function $\psi^{(n)} (x_0)$ of the state $\ket{\psi^{(n)}}_0$ in a similar manner to Ref.~\cite{Takase2021Generation}.
The normalized wave function $\Phi_{\text{Ent}}(x_0, x_1)$ of the two-mode Gaussian entangled state $\ket{\Phi_{\text{Ent}}}_{0,1}$ in Eq.~\eqref{eq:entangledstate_singlemode} is given by~\cite{Takase2021Generation,Tomoda2024Boosting}
\begin{gather}
  \Phi_{\text{Ent}}(x_0, x_1)
  =
  \frac{|\sigma ^{-1}|^{\frac{1}{4}}}{\sqrt{2\pi} }
  \exp
  \left[
    -\frac{1}{4}
    (x_0, x_1) \sigma^{-1} (x_0, x_1) ^\mathsf{T}
  \right]
  ,
\end{gather}
where $\sigma$ is the covariance matrix of the quadrature amplitudes $x_0$ and $x_1$.
The inverse matrix is given by
\begin{gather}
  \sigma^{-1}
  =
  2
  \begin{pmatrix}
    R e^{2r_0} + T e^{2r_1} 
    & \sqrt{RT} (-e^{2r_0} + e^{2r_1}) \\
    \sqrt{RT} (- e^{2r_0} + e^{2r_1} ) 
    & T e^{2r_0} + R e^{2r_1} \\
  \end{pmatrix}
  ,
\end{gather}
where we introduce the BS transmissivity $T\ (=1-R)$ in place of $\kappa_0$ ($T=\sin^2 (\kappa_0/2)$, $R=\cos^2 (\kappa_0/2)$), and the squeezing parameters $r_0$ and $r_1$ are real as noted at the end of Sec.~\ref{sec:TwomodeGaussian}.
The normalized wave function $I^{(n)}(x)$ of the $n$-photon state is given in terms of the $n$-th order Hermite polynomial
$H_n(x)=(-1)^n e^{x^2} \frac{d^n}{dx^n} e^{-x^2}$ as
\begin{gather}
  I^{(n)} (x) 
  =
  \frac{1}{\pi ^{\frac{1}{4}} \sqrt{2^n n!}} H_n(x) e^{-\frac{1}{2} x ^2}
  .
  \label{eq:nphoton}
\end{gather}
Then, the wave function $\psi^{(n)}(x_0)$ of the state $\ket{\psi^{(n)}}_0$ is obtained from Eq.~\eqref{eq:n-photon_projected_normalized} as
\begin{align}
  \psi^{(n)} (x_0)
  = 
  \frac{1}{\sqrt{P(n)}}
  \int dx_1
  I^{(n)} (x_1)
  \Phi_{\text{Ent}}(x_0, x_1).
  \label{eq:psix0_2}
\end{align}
The integrand in Eq.~\eqref{eq:psix0_2} can be written as
\begin{align}
  I^{(n)} (x_1)
  \Phi_{\text{Ent}}(x_0, x_1)
  =
  \underbrace{
    \frac{1}{\pi ^{\frac{1}{4}} \sqrt{2^n n!}}
    \frac{|\sigma ^{-1}|^{\frac{1}{4}}}{\sqrt{2\pi}}
  H_n(x_1)}_
  {\text{(a)}}
  \underbrace{
    e^
    {
      -\frac{1}{4} \sigma^{-1} _{11} x_0 ^2
      -\frac{1}{2} \sigma^{-1} _{12} x_0 x_1
      -\frac{1}{4} \sigma^{-1} _{22} x_1 ^2
      -\frac{1}{2} x_1 ^2
      }
  }_
    {\text{(b)}}
    ,
  \label{eq:IPhi}
\end{align}
where $\sigma_{ij}^{-1}$ denotes the $(i,j)$ element of $\sigma^{-1}$.
The two terms (a) and (b) in Eq.~\eqref{eq:IPhi} are rearranged separately as follows.
First, the exponential term (b) in Eq.~\eqref{eq:IPhi} can be rewritten as
\begin{gather}
  \text{(b)}
  = 
  e^{- (A_1 x_1 - A_2 x_0) ^2 }
  e^{- \frac{1}{2}A_3 ^2  x_0 ^2 }
  ,
  \label{eq:Hn_x1_1}
\end{gather}
where the coefficients $A_1, A_2$, and $A_3$ depend on $(r_0, r_1, T)$ as
\begin{gather}
  \begin{dcases}
   A_1(r_0, r_1, T) = 
   \sqrt{\frac{1}{2} + \frac{1}{4}\sigma_{22} ^{-1}}
   ,
   \\
   A_2(r_0, r_1, T) = 
   -\frac{\sigma_{12} ^{-1}}{4A_1(r_0, r_1, T)} 
  ,
   \\
   A_3(r_0, r_1, T) = 
   \sqrt{\frac{1}{2} \sigma_{11} ^{-1} - \frac{(\sigma_{12} ^{-1})^2}{4+2\sigma_{22}^{-1}}}
   ,
  \end{dcases}
  \label{eq:A1A2A3}
\end{gather}
and all of them are nonzero provided that the two-mode Gaussian state is entangled.
Second, since the $n$-th order Hermite polynomial is a degree-$n$ polynomial, the polynomial term (a) in Eq.~\eqref{eq:IPhi} can be rewritten by introducing coefficients $\{d_{n,m}\}_{m=0}^{n}$ as
\begin{align}
  \text{(a)}
  & =
    \sum_{m=0}^{n}
    d_{n,m} x_1^m
  \label{eq:Hn_x1_3}
  \\
  & =
    \sum_{m=0}^{n}
    d_{n,m}
    \left(
    \frac{1}{A_1}
    \right)^m
    \left[
      (A_1 x_1 - A_2 x_0) + A_2 x_0
    \right]^m
  \label{eq:Hn_x1_4}
  \\
  & =
  \sum_{m=0}^{n}
  \left\{
    d_{n,m}
    \left(
    \frac{1}{A_1}
    \right)^m
    \sum_{l=0} ^{m}
    \left[
      \begin{pmatrix}
        m\\
        l\\
      \end{pmatrix}
      (A_1 x_1 - A_2 x_0) ^{m-l}
      (A_2 x_0) ^l
      \right]
  \right\}
  .
  \label{eq:Hn_x1_5}
\end{align}
From Eqs.~\eqref{eq:Hn_x1_1} and \eqref{eq:Hn_x1_5}, we can rewrite Eq.~\eqref{eq:IPhi} as 
  \begin{align}
    I^{(n)} (x_1)
    \Phi_{\text{Ent}}(x_0, x_1)
    &=
    \sum_{m=0}^{n}
    \left\{
      d_{n,m}
      \left(
        \frac{1}{A_1}
        \right)^m
        \sum_{l=0} ^{m}
        \left[
        \begin{pmatrix}
          m\\
          l\\
        \end{pmatrix}
        (A_2 x_0) ^l
        e^{- \frac{1}{2}A_3 ^2  x_0 ^2 }
        (A_1 x_1 - A_2 x_0) ^{m-l}
        e^{- (A_1 x_1 - A_2 x_0) ^2 }
        \right]
        \right\}
      .
\end{align}
Thus, the wave function $\psi^{(n)} (x_0)$ in Eq.~\eqref{eq:psix0_2} is given by
\begin{align}
  \psi^{(n)} (x_0)
  =
  \frac{1}{\sqrt{P(n)}}
    \sum_{m=0}^{n}
    \left\{
      d_{n,m}
      \left(
        \frac{1}{A_1}
        \right)^m
        \sum_{l=0} ^{m}
        \left[
        \begin{pmatrix}
          m\\
          l\\
        \end{pmatrix}
        \underbrace{
        (A_2 x_0) ^l
        }
        _{\text{(c)}}
        e^{- \frac{1}{2}A_3 ^2  x_0 ^2 }
        \underbrace{
        \int dx_1
        (A_1 x_1 - A_2 x_0) ^{m-l}
        e^{- (A_1 x_1 - A_2 x_0) ^2 }
        }
        _{\text{(d)}}
        \right]
    \right\}
      .
      \label{eq:psix1}
\end{align}
The terms (c) and (d) can be rewritten by introducing coefficients $\{d'_{l,k}\}_{k=0} ^{l}$ and  $d''_{m,l}$ as
\begin{align}
  &
  \text{(c)}
  =
  \sum_{k=0}^{l}
  d'_{l,k} H_k \left(A_3 x_0 \right)
  ,
  \label{eq:hndecomposition}
  \\
  &
  \text{(d)}
  = 
  \frac{1}{A_1}
  \int dx'_1
  {(x'_1)} ^{m-l}
  e^{- {(x'_1)} ^2}
  =
  \frac{1}{A_1}
  d''_{m,l}.
  \label{eq:cintegral}
\end{align}
Then, using Eqs.~\eqref{eq:hndecomposition} and \eqref{eq:cintegral}, Eq.~\eqref{eq:psix1} can be rewritten as
\begin{align}
  \psi^{(n)} (x_0)
  &=
  \frac{1}{\sqrt{P(n)}}
    \sum_{m=0}^{n}
    \left[
      d_{n,m}
      \left(
        \frac{1}{A_1}
        \right)^{m+1}
        \sum_{l=0} ^{m}
        \left\{
        \begin{pmatrix}
          m\\
          l\\
        \end{pmatrix}
        d''_{m,l}
      \sum_{k=0}^{l}
      \left[
      d'_{l,k}
      H_k \left(A_3 x_0 \right)
        e^{- \frac{1}{2}A_3 ^2  x_0 ^2 }
        \right]
      \right\}
    \right].
    \label{eq:psin_includingHn}
  \end{align}
By comparing $H_k \left(A_3 x_0 \right) e^{- \frac{1}{2}A_3 ^2  x_0 ^2 }$ in Eq.~\eqref{eq:psin_includingHn} with the wave function of the $n$-photon state in Eq.~\eqref{eq:nphoton}, we obtain
\begin{align}
  \psi^{(n)} (x_0)
  =
  \sum_{j=0}^{n}
  C''_j {}^{(n)} \left[ \sqrt{A_3} I^{(j)}  \left(A_3 x_0 \right) \right],
  \label{eq:CdashdashIntroduced}
\end{align}
where the coefficients $\{C''_j {}^{(n)}\}_{j=0} ^{n}$ are introduced, and the function $\sqrt{A_3} I^{(j)}  \left(A_3 x_0 \right)$ represents the normalized wave function of the squeezed $j$-photon state.
Therefore, the wave function $\psi^{(n)} (x_0)$ represents a superposition of squeezed $j$-photon states ($j=0,1,\dots,n$). Returning to the bra-ket notation, the corresponding state $\ket{\psi^{(n)}}_0$ is written as
\begin{align}
  \ket{\psi^{(n)}}_0 =
  \hat{S}_0 (r_{\text{out}})
  \left(
  \sum_{j=0} ^{n} 
  C''_j {}^{(n)} \ket{j}_0
  \right).
\end{align}
We here define the squeezing level in the quadrature $x$ as $e^{-2r_{\text{out}}}$.
This squeezing level depends only on the parameters $(r_0, r_1, T)$ used to prepare the two-mode Gaussian entangled state, and is independent of the detected photon number~$n$:
\begin{align}
  e^{-2r_{\text{out}}} &= \frac{1}{A_3 ^2}
  = \frac{1}{\frac{1}{2}\sigma^{-1} _{11} - \frac{(\sigma^{-1} _{12}) ^2}{4+2\sigma^{-1} _{22}}}
  =
  e^{-2r_0 - 2r_1}
  \frac{1+Te^{2r_0} + Re^{2r_1}}{1+Te^{-2r_0} + Re^{-2r_1}}
  .
  \label{eq:rout_formula}
\end{align}

Finally, by superposing all the states $\{\ket{\psi^{(n)}}_0\}_{n=0}^{N}$ as in Eq.~\eqref{eq:superposition}, we obtain the final state $\ket{\psi}_0$ as
\begin{align}
  \ket{\psi}_0
  & \propto
  \sum_{n=0} ^{N}
  C'_n \sqrt{P(n)}
  \left(
  \hat{S}_0 (r_{\text{out}})
  \sum_{j=0} ^{n} C''_j {}^{(n)}  \ket{j}_0
  \right)
  \\
  &=
  \hat{S}_0 (r_{\text{out}})
  \left[
  \sum_{n=0} ^{N}
  C'_n \sqrt{P(n)}
  \left(
  \sum_{j=0} ^{n} 
  C''_j {}^{(n)} \ket{j}_0
  \right)
  \right]
  \label{eq:psi0_pre}
  .
\end{align}
Using coefficients $\{C_n\}_{n=0} ^{N}$ satisfying the following equation,
  \begin{gather}
    C'_0 \sqrt{P(0)}
    \begin{pmatrix}
      C''_0 {}^{(0)} \\
      0 \\
      0 \\
      \vdots \\
      0 \\
    \end{pmatrix}
    +
    C'_1 \sqrt{P(1)}
    \begin{pmatrix}
      C''_0 {}^{(1)} \\
      C''_1 {}^{(1)} \\
      0 \\
      \vdots \\
      0 \\
    \end{pmatrix}
    +
    \cdots
    +
    C'_N \sqrt{P(N)}
    \begin{pmatrix}
      C''_0 {}^{(N)} \\
      C''_1 {}^{(N)} \\
      C''_2 {}^{(N)} \\
      \vdots \\
      C''_N {}^{(N)} \\
    \end{pmatrix}
    =
    \begin{pmatrix}
      C_0 \\
      C_1 \\
      C_2 \\
      \vdots \\
      C_N \\    
    \end{pmatrix}
    ,
    \label{eq:relationformula}
  \end{gather}
Eq.~\eqref{eq:psi0_pre} can be expressed in a simple form as
\begin{align}
  \ket{\psi}_0 \propto
  \hat{S}_0 (r_{\text{out}})
  \sum_{n=0} ^{N} C_n \ket{n}_0.
  \label{eq:psi0_final}
\end{align}
From the above, it has been shown that the final heralded state can always be expressed in the form of Eq.~\eqref{eq:psi0_final}.

We show below that the state $\ket{\psi}_0$ in Eq.~\eqref{eq:psi0_final} is a stellar-rank-$N$ state by proving that $C_N \neq 0$.
From Eq.~\eqref{eq:relationformula}, the coefficient $C_N$ is specifically given by
\begin{align}
  C_N = 
  C'_N 
  C''_N {}^{(N)}
  \sqrt{P(N)}
  .
  \label{eq:C_dash}
\end{align}
As noted after Eq.~\eqref{eq:superposition}, since $P(N) \neq 0$ and $C'_N \neq 0$, it suffices to show that $C''^{(n)}_n \neq 0$ for all $n=0,1,\dots,N$, which includes the specific case $C''_N{}^{(N)} \neq 0$.
By comparing the $H_n(A_3 x_0)$ term in Eq.~\eqref{eq:psin_includingHn} with the $I^{(n)}(A_3 x_0)$ term in Eq.~\eqref{eq:CdashdashIntroduced}, we obtain the following for $k=l=m=n$:
\begin{align}
  C''_n {}^{(n)} 
  =
  d_{n,n}
  d'_{n,n}
  d''_{n,n}
  \pi ^{\frac{1}{4}}
  \sqrt{
  \frac{2^{n} n!}{A_3 P(n)}
  }
  \left( \frac{1}{A_1} \right)^{n+1}
  .
  \label{eq:Cnn}
\end{align}
Here the coefficients $d_{n,n}$, $d'_{n,n}$, and $d''_{n,n}$ are also nonzero, as shown below:
\begin{itemize}
  \item Since $d_{n,n}$ is the coefficient of $x_1^n$ in $H_n(x_1)$, as introduced in Eq.~\eqref{eq:Hn_x1_3}, we have $d_{n,n} \neq 0$.
  \item In Eq.~\eqref{eq:hndecomposition}, $d'_{n,n}$ is the coefficient of $H_n(A_3 x_0)$ in the expansion of $(A_2 x_0)^n$ in terms of Hermite polynomials, and thus $d'_{n,n} \neq 0$.
  \item The coefficient $d''_{n,n}$ is obtained from its definition in Eq.~\eqref{eq:cintegral} as follows, and hence $d''_{n,n} \neq 0$.
  \begin{align}
    d''_{n,n}
    =
    \int dx'_1
    e^{- {(x'_1)} ^2}
    \neq 0
  \end{align}
\end{itemize}
Consequently, the coefficients $\{C''^{(n)}_n\}_{n=0}^{N}$ in Eq.~\eqref{eq:Cnn} are nonzero.
The condition $C''^{(N)}_N \neq 0$ guarantees that $C_N \neq 0$ via Eq.~\eqref{eq:C_dash}.
Therefore, the state $\ket{\psi}_0$ is shown to be a stellar-rank-$N$ state.

\subsubsection{How to set experimental parameters for target state}
In this section, we prove that our light source can generate any target state of the form given in Eq.~\eqref{eq:psi0_final}. 
To this end, we describe below how to choose the experimental parameters $(r_0, r_1, T)$ and $\{\alpha_j\}_{j=1}^{N}$ 
so as to generate an arbitrary target state.

\begin{enumerate}
  \item We consider the generation of a target state of the form given in Eq.~\eqref{eq:psi0_final}, characterized by the parameters $r_{\text{out}} \in \mathbb{R}$ and $C_0, C_1, \dots, C_N \in \mathbb{C} \ (C_N \neq 0)$.
  \item We choose the parameters $(r_0, r_1, T)$ to achieve the target squeezing level $e^{-2r_{\text{out}}}$ based on Eq.~\eqref{eq:rout_formula}. Any value of $r_{\text{out}}$ can be realized by an appropriate choice of $(r_0, r_1, T)$. At the end of this section, we mention the remaining flexibility in $(r_0, r_1, T)$ under Eq.~\eqref{eq:rout_formula}.
  \item Once $(r_0, r_1, T)$ are fixed, the state $\ket{\psi^{(n)}}_0$ heralded by $n$-photon detection and the corresponding probability $P(n)$ are determined from Eqs.~\eqref{eq:n-photon_projected_normalized} and \eqref{eq:eq:n-photon_probability}, thereby determining the coefficients $\{C''_j {}^{(n)}\}_{0 \leq j \leq n \leq N}$ and $\{P(n)\}_{n=0} ^{N}$.
  \item Equation~\eqref{eq:relationformula} relates the following quantities:
  \begin{itemize}
    \item $\{C'_n\}_{n=0} ^{N}$ (to be determined, tunable via the experimental parameters $\{\alpha_j\}_{j=1} ^{N}$)
    \item $\{C_n\}_{n=0} ^{N}$ (specified by the target state)
    \item $\{C''_j {}^{(n)}\}_{0 \leq j \leq n \leq N}$ and $\{P(n)\}_{n=0} ^{N}$ (determined in the previous step)
  \end{itemize}
In Eq.~\eqref{eq:relationformula}, using the fact that $\{P(n)\}_{n=0} ^{N}$ and $\{C''_n {}^{(n)}\}_{n=0} ^{N}$ are nonzero, as discussed in Sec.~\ref{sec:statetype}, the coefficients $\{C'_n\}_{n=0} ^{N}$ can be uniquely determined in the order $C'_N, C'_{N-1}, \dots, C'_0$.
  \item From Eqs.~\eqref{eq:finalprojector_pre} and \eqref{eq:finalprojector}, the coefficients $\{C'_n\}_{n=0}^{N}$ are related to the displacement amplitudes $\{\alpha_j\}_{j=1}^{N}$ as follows:
  \begin{align}
  &
  \bigotimes_{j=1} ^{N}
  \left(
    \frac{1}{\sqrt{N}c'}
    \ann{1} [f_1]
    + \alpha_j
  \right)
  \propto
  \sum_{n=0} ^{N}
  \frac{C'_n}{\sqrt{n!}}
  \left(\ann{1} [f_1] \right) ^n
  .
  \label{eq:alpha_derivation}
  \end{align}
By replacing the operator $\ann{1} [f_1]$ in Eq.~\eqref{eq:alpha_derivation} with a complex variable, the left-hand side becomes a product of $N$ linear terms, while the right-hand side represents an $N$-th degree polynomial.  
According to the fundamental theorem of algebra, any $N$-th degree polynomial can be uniquely factorized into a product of $N$ linear terms.  
Therefore, for any given set of coefficients  $\{C'_n\}_{n=0}^{N}$, one can always find a corresponding set of parameters $\{\alpha_j\}_{j=1}^{N}$ that satisfies Eq.~\eqref{eq:alpha_derivation}.  
This approach is analogous to that in Ref.~\cite{Yukawa2013Generating}.
\end{enumerate}

Thus, we have shown that any target state in Eq.~\eqref{eq:psi0_final} can be generated by appropriately setting the parameters $(r_0, r_1, T)$ and $\{\alpha_j\}_{j=1} ^{N}$ following the above steps.  
Crucially, the initial selection of $(r_0, r_1, T)$ uniquely determines the remaining parameters.  
As noted above, some flexibility remains in the choice of $(r_0, r_1, T)$.  
In general, it is preferable to choose $(r_0, r_1, T)$ such that the mean photon number in the trigger path is relatively small, which in turn keeps the displacement amplitudes $\{\alpha_j\}_{j=1} ^{N}$ relatively small.  
In this regime, we can satisfy the assumption as noted before Eq.~\eqref{eq:projector} in Sec.~\ref{sec:Heralding non-Gaussian states in the target TW}, which requires that each displacement amplitude is small ($|\alpha_j|\ll 1$) and that multiphoton detections at each on/off detector are negligible.
On the other hand, reducing the mean photon number in the trigger path lowers the generation probability.  
Therefore, in practice, the choice of $(r_0, r_1, T)$ represents a trade-off between the generation rate and the state quality.

\section{Experimental details}
\subsection{Setup details}

\begin{figure}[!htbp]
  \centering
  \includegraphics[width=0.99\linewidth]{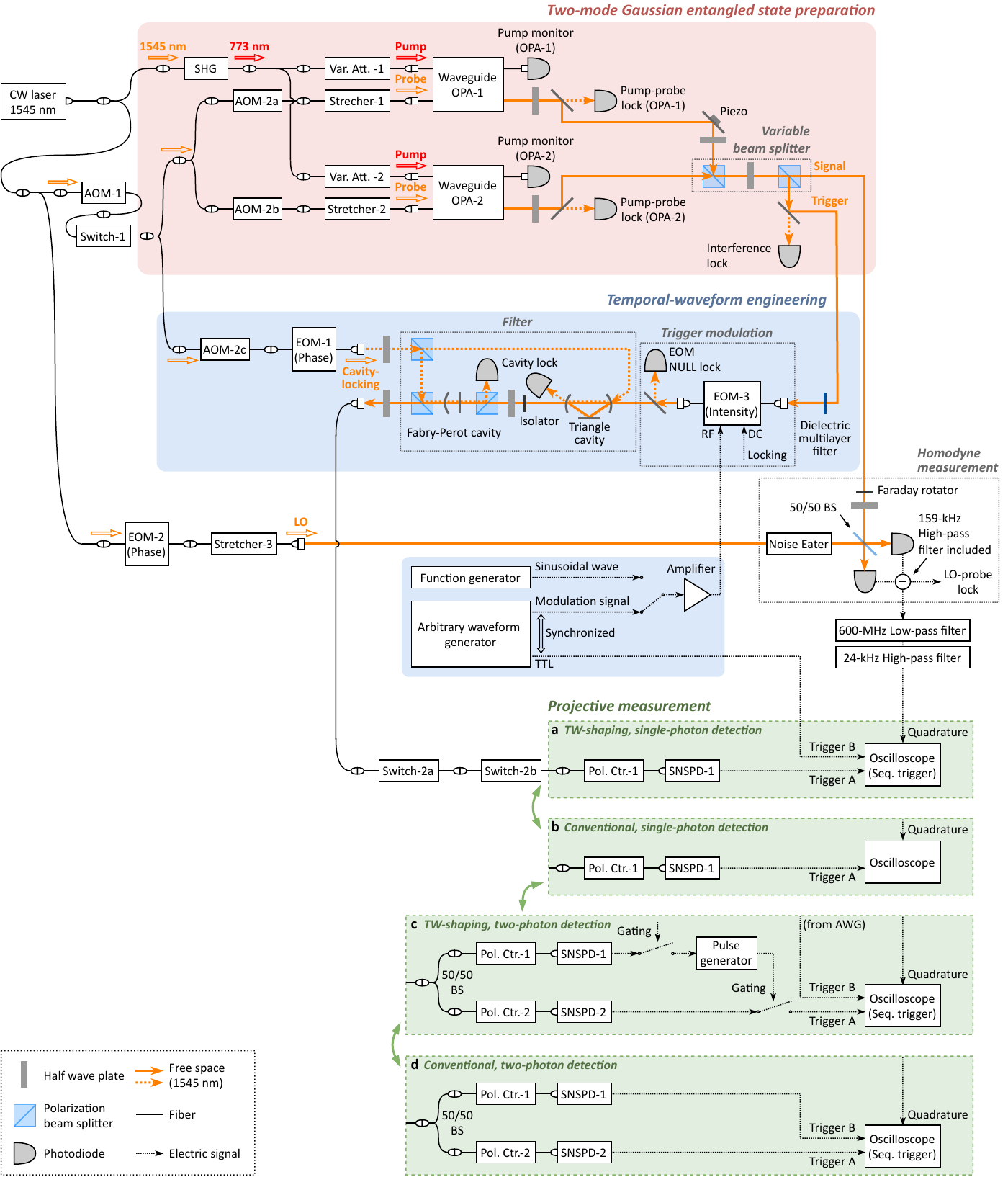}
  \caption{
  Experimental setup.
  CW, continuous wave;
  SHG, second harmonic generator;
  AOM, acousto-optic modulator;
  Var. Att., variable attenuator;
  OPA, optical parametric amplifier;
  EOM, electro-optic modulator;
  SNSPD, superconducting nanostrip single-photon detector;
  LO, local oscillator;
  BS, beam splitter;
  Pol. Ctr., polarization controller;
  Seq. trigger, sequence trigger;
  TW, temporal waveform.
  The green areas indicate the trigger systems for four cases:
  a.~TW-shaping case with single-photon detection,
  b.~conventional case (without TW-shaping) with single-photon detection,
  c.~TW-shaping case with two-photon detection, and
  d.~conventional case (without TW-shaping) with two-photon detection.
  \label{fig:Sup_setup}
  }
\end{figure}

Our detailed setup is shown in Fig.~\ref{fig:Sup_setup}.  
The basic configuration for non-Gaussian state generation, including entangled state preparation and trigger-light filtering, is the same as in our previous experiment~\cite{Tomoda2024Boosting}.  
Here, we focus on describing the newly constructed part.

\begin{figure}[!h]
  \centering
  \includegraphics[width=0.95\linewidth]{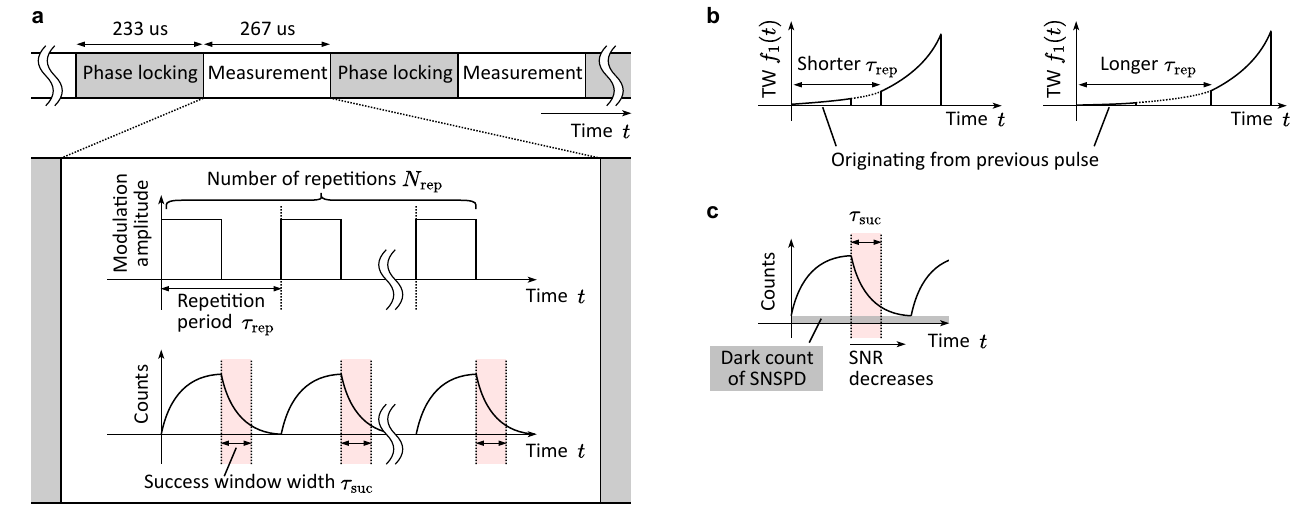}
  \caption{Parameters related to temporal-waveform (TW) engineering.
  All figures present the case of square-pulse modulation as an example.
  a.~Experimental sequence and definitions of TW-engineering parameters.
  b.~TW of heralded states depending on modulation repetition period $\tau_{\text{rep}}$.
  c.~Effect of SNSPD dark counts.
  \label{fig:Sup_seq}
  }
\end{figure}

In the trigger path (optical setup highlighted in blue in Fig.~\ref{fig:Sup_setup}), between the dielectric multilayer filter and two cavities, a fiber-coupled electro-optic intensity modulator (Exail, MXER-LN-10-00-P-P-FA-FA-40dB; bandwidth \SI{10}{\giga\hertz}, extinction ratio \SI{40}{\deci\bel}) is installed for TW control.  
This modulator, labeled EOM-3 in Fig.~\ref{fig:Sup_setup}, has two electrical input ports: a \textit{DC port}, used to compensate the drift of two arm lengths of the internal Mach-Zehnder interferometer, and an \textit{RF port}, used for high-speed modulation of the incoming light.
Our system alternately repeats the phase-locking control and quantum-state measurement at \SI{2}{\kilo\hertz} as in the previous experiment~\cite{Tomoda2024Boosting}, as shown in Fig.~\ref{fig:Sup_seq}a.
As for the newly-introduced EOM-3, we lock its transmissivity to null during the control period.
For this locking, a sinusoidal signal is applied to the RF port, and a fraction of the output light is detected.
The detection signal is demodulated with the same sinusoidal reference to generate an error signal, which is fed back to the DC port for locking.
During the measurement period, in addition to the DC-port control signal, a modulation signal for TW control is applied to the RF port from an arbitrary waveform generator (AWG; Keysight 81160A; bandwidth \SI{330}{\mega\hertz}), as illustrated in Fig.~\ref{fig:Sup_seq}a.

Our TW shaping is successful only when photon detection occurs within a specific time window.  
For single-photon detection, this determination is performed using the sequence-trigger function of an oscilloscope (Tektronix, MSO56; bandwidth \SI{1}{\giga\hertz}), as illustrated in Fig.~\ref{fig:Sup_setup}a.  
In this function, data acquisition by Trigger-B occurs only if the second trigger (Trigger-B) arrives within a defined time window after the first trigger (Trigger-A).  
We use the photon-detection signal from the superconducting nanostrip single-photon detector (SNSPD) as Trigger-A and the modulation-synchronized TTL signal from the AWG as Trigger-B, with the time window set appropriately as described later.

While this sequence trigger alone is sufficient for single-photon detection, two-photon detection requires more complex conditioning.  
As shown in Fig.~\ref{fig:Sup_setup}c, the trigger light after Switch-2b is split into two paths and detected by two independent SNSPDs.  
To ensure successful TW shaping, each detection signal must satisfy the time-window condition, which is implemented in the following manner:  
First, the signal from SNSPD-1 is sent to a gate switch that is enabled only during the success time window.  
This gated signal then triggers a pulse generator, which outputs a square pulse to gate the SNSPD-2 signals.  
This arrangement ensures that only SNSPD-2 signals associated with the same modulation pulse as the SNSPD-1 trigger are sent to the oscilloscope.  
These signals are then used as Trigger-A for the sequence trigger, allowing us to verify whether the SNSPD-2 signals also fall within the success window.

We note that a different configuration is used for the conventional case without modulation.  
For single-photon detection, the oscilloscope is triggered directly by the photon-detection signal (Fig.~\ref{fig:Sup_setup}b).  
For two-photon detection, the two detection events must occur simultaneously; in practice, we allow a timing difference of up to \SI{5}{\nano\second}.  
This condition is implemented using the sequence-trigger function, with the two detection signals serving as Trigger-A and Trigger-B (Fig.~\ref{fig:Sup_setup}d).

Additionally, the homodyne-measurement setup used in our previous experiment~\cite{Tomoda2024Boosting} is slightly modified.
A noise eater is installed in the path of the local-oscillator (LO) light to stabilize the LO power during long-term measurements, as required for two-photon state generation.  
Furthermore, a combination of a Faraday rotator and a half-wave plate (HWP) is inserted in the signal path to suppress fake photon-detection triggers caused by back-reflection of the LO light at the homodyne detector.

\subsection{Experimental parameters}
Here, we describe the parameters used to obtain the results shown in Fig.~4 of the main text.  

For state engineering, to generate single-photon, Schr\"odinger cat, and two-photon states, we adjusted the initial two-mode Gaussian entangled state by setting the two input squeezing levels, $e^{-2r_0}$ and $e^{-2r_1}$, and the transmissivity $T$ of the variable BS, as summarized in Table~\ref{table:Parameter settings}.  
Among the various combinations of $(r_0, r_1, T)$ capable of generating the same target state, we selected the specific values listed in Table~\ref{table:Parameter settings}, taking into account the trade-off between the generation rate and the state quality caused by multiphoton detection at the on/off detector.

\begin{table}[!hbtp]
    \caption{State-engineering parameters}
    \label{table:Parameter settings}
    \centering
      \begin{tabular*}{0.55\columnwidth}{l@{\extracolsep{\fill}}ccc}
        \toprule
        & \multicolumn{2}{c}{Input squeezing levels} & {Transmissivity} \\
        \cmidrule{2-3} \cmidrule{4-4}
        State type & \quad $e^{-2r_0}$ & $e^{-2r_1}$ & $T$ \\
        \midrule
        Single-photon & \quad \SI{3}{\deci\bel} & \SI{-3}{\deci\bel} & 0.50  \\
        Cat           & \quad \SI{5}{\deci\bel} & \SI{-1}{\deci\bel} & 0.14  \\
        Two-photon    & \quad \SI{4}{\deci\bel} & \SI{-4}{\deci\bel} & 0.50  \\
        \bottomrule
    \end{tabular*}
\end{table}

\begin{figure}[!h]
\centering
\includegraphics[width=0.6\linewidth]{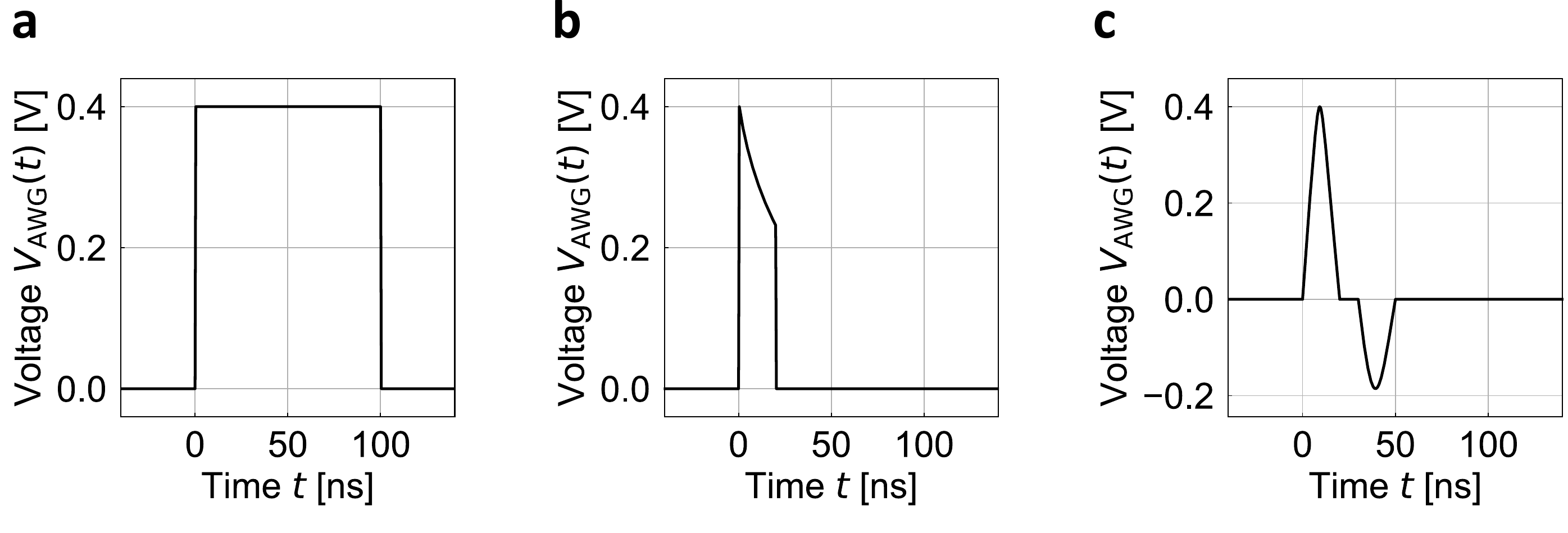}
\caption{
  Arbitrary waveform generator (AWG) setting voltage for three modulation cases: a.~square-pulse modulated waveform, b.~square-pulse waveform, and c.~balanced time-bin waveform.
\label{fig:AWG}
}
\end{figure}

For TW engineering via trigger-light modulation, we set the AWG voltage as shown in Fig.~\ref{fig:AWG}.
The AWG voltage $V_{\text{AWG}}(t)$ and the final TW $f_1(t)$ of the heralded state in Eq.~\eqref{eq:relationfandm} is related as 
\begin{align}
  f_1(t) \propto e^{\gamma t}
  \sin \left[
    \frac{\pi}{2} \frac{V_{\text{AWG}}(t)}{V_0}
    \right],
\end{align}
where the constant $V_0$ was estimated from a preliminary calibration experiment, $V_0 = \SI{0.5}{\volt}$, and $V_{\text{AWG}}(t)$ is set within the range from \SI{-0.4}{\volt} to \SI{0.4}{\volt} in each experiment.
Except for the square-pulse modulated waveform, we compute $V_{\text{AWG}}(t)$ from the above relation using $\gamma = 2\pi \times \SI{2.9}{\mega\hertz}$, and the prepared waveform data is transferred to the AWG via the dedicated control software.
In the measurement period, each modulation is repeated as shown in Fig.~\ref{fig:Sup_seq}a.
The repetition period $\tau_{\text{rep}}$ and the number $N_{\text{rep}}$ of repetitions are chosen to the values in Table~\ref{table:TW-engineering parameters}.
We determine the values of $\tau_{\text{rep}}$ to sufficiently suppress the effect of previous modulation pulses (to less than 1\%).
Figure~\ref{fig:Sup_seq}b indicates that a shorter $\tau_{\text{rep}}$ causes the TW to deviate from the ideal profile due to the previous pulses.

In our TW-engineering scheme, the trigger light must be detected after the modulation has ended (Fig.~\ref{fig:S2}).
However, the criteria for determining the end of the success window has yet to be discussed.
We clarify this point by considering the influence of fake photon-detection triggers arising from the dark counts of the SNSPDs.
Figure~\ref{fig:Sup_seq}c illustrates that trigger-light detection events follow a temporal distribution governed by the modulation function (square-pulse modulation case in this figure) and cavity response, whereas the dark counts occur stochastically, independent of the modulation timing.
Consequently, the signal-to-noise ratio (SNR) is gradually degraded after the modulation ends (\textit{signal}: trigger-light detection events; \textit{noise}: dark counts).
The success window width $\tau_{\text{suc}}$ should be set to exclude the regions with poor SNR.
Since the detection counts of trigger photons depend on the specific state-engineering parameters, $\tau_{\text{suc}}$ is determined individually for each state, as summarized in Table~\ref{table:TW-engineering parameters}.
For the two-photon state generation with the balanced time-bin waveform, we set $\tau_{\text{suc}}$ to a relatively longer duration to ensure a practical generation rate, while accepting a slight reduction in the state quality.

\begin{table}[h]
    \caption{Temporal-waveform engineering parameters}
    \label{table:TW-engineering parameters}
    \centering
    \begin{tabular*}{0.8\columnwidth}{@{\extracolsep{\fill}}lccc}
        \toprule
        & Square-pulse modulated & Square-pulse & Balanced time-bin \\
        \midrule
        Repetition period ($\tau_{\text{rep}}$) & \SI{200}{\nano\second} & \SI{220}{\nano\second} & \SI{150}{\nano\second} \\
        \addlinespace[1ex] 
        Number of repetitions ($N_{\text{rep}}$) & 1200 & 1100 & 1600 \\
        \addlinespace[1ex] 
        Success window width ($\tau_{\text{suc}}$) \\
        \quad Single-photon & \SI{80}{\nano\second} & \SI{50}{\nano\second} & \SI{30}{\nano\second} \\
        \quad Cat           & \SI{60}{\nano\second} & \SI{30}{\nano\second} & \SI{10}{\nano\second} \\
        \quad Two-photon    & \SI{80}{\nano\second} & \SI{50}{\nano\second} & \SI{40}{\nano\second} \\
        \bottomrule
    \end{tabular*}
\end{table}

\subsection{Measurement \& Analysis}
In each experiment, to estimate the TW of the heralded state and evaluate its quality, we measure time-series quadrature amplitudes at 12 LO phases, incremented by \SI{15}{\degree}.
The quadrature data are obtained by a homodyne detector (bandwidth \SI{200}{\mega\hertz}) and digitized using an oscilloscope (bandwidth \SI{1}{\giga\hertz}).
Data are acquired in \SI{2}{\micro\second}-long frames at a sampling rate of \SI{3.125}{\giga\sample \per \second}.
For the single-photon and cat states, \num{5000} data frames are acquired at each phase.
For the two-photon states, \num{2000}, \num{2000}, \num{1000}, and \num{500} frames are acquired for the cases of the exponentially rising waveform (conventional scheme), square-pulse modulated waveform, square-pulse waveform, and balanced time-bin waveform, respectively.
During the acquisition, a portion of the time stamps is also recorded, from which the generation rates are calculated, as listed in Table~\ref{table:generation rate} with standard errors.

\begin{table}[h]
    \caption{Generation rate (counts/s)}
    \label{table:generation rate}
    \centering
    \begin{tabular*}{\columnwidth}{@{\extracolsep{\fill}} lcccc}
        \toprule 
        & \multicolumn{1}{c}{Conventional scheme} & \multicolumn{3}{c}{Our scheme} \\
        \cmidrule{2-2} \cmidrule{3-5} 
        & Exponentially rising & Square-pulse modulated & Square-pulse & Balanced time-bin \\
        \midrule 
        Single-photon & $(4.1 \pm 0.5) \times 10^{4}$ & $(4.0 \pm 0.4) \times 10^{3}$ & $(1.3 \pm 0.2) \times 10^{3}$ & $(6.4 \pm 0.8) \times 10^{2}$\\
        Cat           & $(2.6 \pm 0.4) \times 10^{4}$ & $(2.7 \pm 0.4) \times 10^{3}$ & $(6 \pm 1) \times 10^{2}$     & $(1.8 \pm 0.1) \times 10^{2}$\\
        Two-photon    & $(2.1 \pm 0.3) \times 10^{1}$ & $(1.0 \pm 0.1) \times 10^{1}$ & $(9 \pm 2) \times 10^{-1}$    & $(1.8 \pm 0.3) \times 10^{-1}$\\
        \bottomrule 
    \end{tabular*}
\end{table}

We perform principal component analysis (PCA) to estimate the TW of the heralded state, following Ref.~\cite{Morin2013Experimentally}.
As a preliminary step, we remove electric circuit noise around $\sim$\SI{160}{\mega\hertz} from the time-series quadrature data by applying a digital finite-impulse-response low-pass filter with 100-MHz cutoff.
We perform PCA on the filtered data combined across different LO phases, and obtain the eigenvectors (TWs) which are denoted as the first, second, and subsequent principal components (PC-1, PC-2, \dots) in descending order of their eigenvalues (quadrature variances).

\begin{figure}[b]
  \centering
  \includegraphics[width=\linewidth]{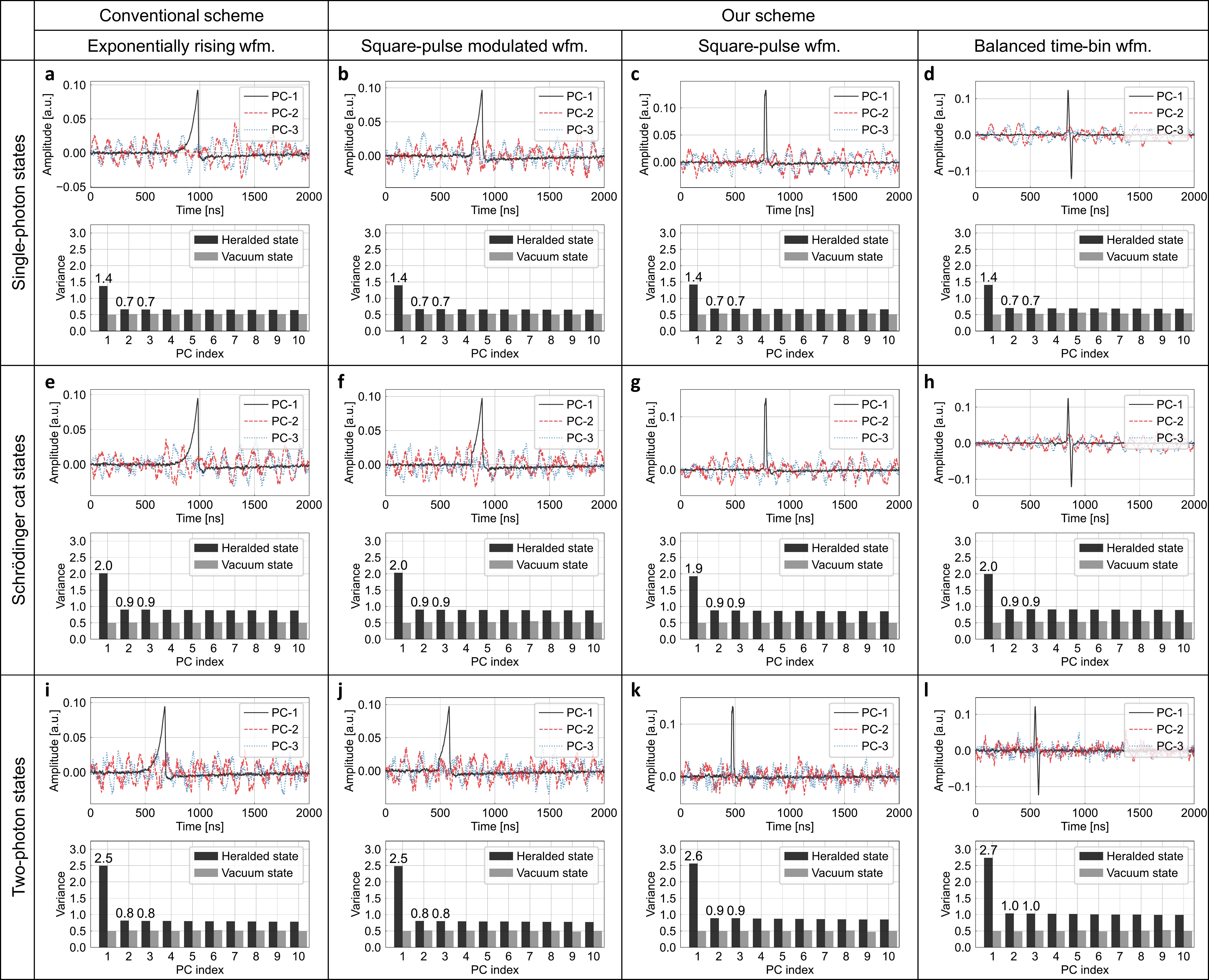}
  \caption{
    Detailed results of the principal component analysis.
    wfm. indicates waveform.
    In each panel, the upper figure illustrates the temporal waveforms (TWs) derived as the first three principal components (PC-1, PC-2, and PC-3), and the lower figure shows the corresponding quadrature variance for each TW.
    The TWs are shown for a \SI{2}{\micro\second} duration, corresponding to the length of the recorded time-series quadrature data.
  \label{fig:S4}
  }
\end{figure}

Figure~\ref{fig:S4} shows the obtained TWs and the corresponding quadrature variances.
The following procedure is used to obtain these results.
First, the TWs (PC-1, PC-2, \dots) and variances are calculated for the data of the heralded states.
Second, we measure the quadrature amplitude of vacuum states by blocking the signal light to the homodyne detector, and calculate the quadrature variance for the same TWs (PC-1, PC-2, \dots).
Then, the variances for both the heralded state and vacuum state data are normalized so that the vacuum-noise variance for PC-1 is 0.5 ($\hbar=1$).
In Fig.~\ref{fig:S4}, the vacuum-noise variances are almost equivalent for all the principal components, which indicates that the quadrature data does not include any large-variance components such as electric circuit noise.
In contrast, for the heralded states, only PC-1 exhibits a larger variance than the other principal components, and these subsequent components (PC-2, PC-3, \dots) show almost equivalent values in each experiment.
Hence, we confirm that PC-1 corresponds to the TW of the heralded states.

We perform tomography to evaluate the quality of the heralded state in the target TW.
We start from calculating the quadrature amplitude for the wave-packet mode by using the theoretical TW which is plotted as a black dotted line in Fig.~4 in the main text.
Next, from these quadratures, we reconstruct the density matrices of the states using the maximum likelihood estimation method~\cite{Lvovsky2004Iterative}, where we set a photon-number cutoff to ten.
We then compute the corresponding Wigner functions and check their minimum values as a non-classicality indicator.
These values are shown in Fig.~4 in the main text with the error bars, which are calculated based on Fisher information matrix~\cite{Rehacek2008Tomography}.
As mentioned in the main text, their qualities can be consistently explained by the losses within our experimental setup and the TW mismatch (between the actual and theoretical waveforms).
Table~\ref{table:signal} shows a detailed breakdown of these losses, based on typical values for the conventional exponentially rising waveform.

\begin{table}[hbtp]
  \caption{Loss breakdown (typical values for the exponentially rising waveform)}
  \label{table:signal}
  \centering
  \begin{tabular*}{0.88\columnwidth}{@{\extracolsep{\fill}} lr}
    \toprule
    Item  &  Typical value\\
    \midrule
    Internal loss of waveguide optical parametric amplifier (OPA) modules & 9\%\\
    Propagation loss in signal path& 8\%\\
    Spatial mode mismatch & 8\%\\
    Inefficiency of photodiodes  &  1\%\\
    Circuit noise of homodyne detector & 2\%\\
    Fake photon-detection triggers of superconducting nanostrip single-photon detector (SNSPD) & 1\%\\
    Temporal-waveform (TW) mismatch & 9\%\\
    \midrule
    Total & 33\%
    \\
    \bottomrule
  \end{tabular*}  
\end{table}

\bibliography{SupplementaryInformation}
